\documentclass{JHEP3}

\usepackage{amssymb}
\usepackage{amsfonts}
\usepackage{amsbsy}
\usepackage{amsmath}
\usepackage{amsthm}
\usepackage{graphicx}
\usepackage{epstopdf}
\usepackage[vcentermath]{youngtab}
\usepackage{multirow}
\usepackage{latexsym}
\usepackage{array}

\def\Z{\mathbb{Z}}

\def\C{\mathbb{C}}

\def\P{\mathbb{P}}

\def\n3a{t}

\def\tr{{\mathrm{tr}}}

\def\phiorig{\phi}
\def\phit{\phi_0}
\def\phif{\phi_1}
\def\phifb{\nu}
\def\phis{\psi_2}
\def\phisa{\phi_2}
\def\phie{\psi_1}
\def\ola{\psi_3}
\def\alphabeta{\xi}
\def\pz{\phit}

\def\betah{\rho}

\def\ho{h^{1, 1}}
\newcommand{\eq}[1]{(\ref{#1})}

\title{Matter and singularities}

\author{David R.  Morrison$^{1,2}$ and Washington Taylor$^3$\\
$^1$Departments of Mathematics and  
Physics\\ University of California, Santa Barbara\\ Santa Barbara, CA 93106, USA\\
\\
$^2$Institute for the Physics and Mathematics of the Universe\\
University of Tokyo\\
Kashiwa, Chiba 277-8582, Japan\\
\\
$^3$Center for Theoretical Physics\\
Department of Physics\\
Massachusetts Institute of Technology\\
77 Massachusetts Avenue\\
Cambridge, MA 02139, USA\\
\\
{\tt drm} {\rm at} {\tt math.ucsb.edu},
{\tt wati} {\rm at} {\tt mit.edu}
}

\preprint{UCSB Math 2011-10, IPMU11-0108, MIT-CTP-4200}

\abstract{We analyze the structure of matter representations arising
  from codimension two singularities in F-theory, focusing on gauge
  groups $SU(N)$.  We give a detailed local description of the
  geometry associated with several types of singularities and the
  associated matter representations.  We also construct global
  F-theory models for 6D and 4D theories containing these matter
  representations.  The codimension two singularities encountered
  include examples where the apparent Kodaira singularity type does
  not need to be completely resolved to produce a smooth Calabi-Yau,
  examples with rank enhancement by more than one, and examples where
  the 7-brane configuration is singular.  We identify novel phase
  transitions, in some of which the gauge group remains fixed but the
  singularity type and associated matter content change along a
  continuous family of theories.  Global analysis of 6D theories on
  $\P^2$ with 7-branes wrapped on curves of small degree reproduces
  the range of 6D supergravity theories identified through anomaly
  cancellation and other consistency conditions.  Analogous 4D models
  are constructed through global F-theory compactifications on $\P^3$,
  and have a similar pattern of $SU(N)$ matter content.  This leads to
  a constraint on the matter content of a limited class of 4D
  supergravity theories containing $SU(N)$ as a local factor of the
gauge group.}

\begin{document}

\section{Introduction}

Over the last decade, the development of D-branes in string theory has
led to dramatic new insights into the connection between gauge theory
and geometry.  This connection is made particularly explicit in the
language of F-theory \cite{Vafa-f, Morrison-Vafa, Morrison-Vafa-II},
where  gauge theory coupled to supergravity
in an even  number of space-time dimensions is described by an elliptically
fibered Calabi-Yau manifold over a base $B$ of complex dimension $d$
for a low-energy theory in $10-2d$ space-time dimensions.
Recent reviews of the aspects of F-theory relevant for the discussion
in this paper are given in \cite{Denef-F-theory, WT-TASI}.

In F-theory, the structure of the gauge group in the low-energy theory
is primarily encoded in the singularities of the elliptic fibration
(with certain global aspects of the gauge group encoded in the
Mordell--Weil and Tate--Shafarevich groups of the elliptic fibration
\cite{pioneG,triples}).  In the
language of type IIB string theory, the gauge group is carried by
7-branes wrapped on topologically nontrivial cycles (divisors) of the
F-theory base manifold $B$.  In the geometrical language of F-theory
such 7-branes are characterized by complex codimension one
singularities in the structure of the elliptic fibration.  Such
codimension one singularities were systematically analyzed by Kodaira
\cite{Kodaira} well before the advent of F-theory.  For a base of
complex dimension one, such singularities are characterized by the
familiar ADE classification of simple Lie algebras.  For each type
of codimension one singularity, the low-energy gauge group contains a
local factor with the associated nonabelian Lie algebra.  When the base is of
higher dimension, monodromies around these codimension one loci can
give rise to non-simply laced groups as well as the simply-laced
groups found on bases of dimension one \cite{Bershadsky-all}.

While the geometry of gauge groups is well understood in F-theory, the
geometry of matter representations in such theories has only been
worked out in a limited set of cases, and there is no general
classification of the range of possibilities.  Many types of matter
representations can arise from local codimension two singularities in
the elliptic fibration in the F-theory picture.  Other types of matter
(such as matter in the adjoint representation of $SU(N)$) can arise
from the global structure of the divisor locus \cite{enhanced,Witten-mf}.  For
the simplest types of representations, such as the fundamental
representation of ADE groups, or the two-index antisymmetric
representation of $SU(N)$, matter fields arise from a local rank one
enhancement of the singularity structure, and the matter content is
easily determined from a decomposition of the adjoint representation
of the correspondingly enhanced group, as described by Katz and Vafa
\cite{Katz-Vafa}.  When the singularity structure of the elliptic
fibration becomes more intricate, the associated matter representations
become more exotic.  Other examples associated with rank one
enhancement were worked out in \cite{Bershadsky-all, Katz-Vafa, Grassi-Morrison,
  Grassi-Morrison-2}.  In this paper we consider rank one enhancements
as well as
other kinds of
singularity structures.  In some cases the apparent Kodaira
singularity associated with a coordinate transverse to the brane does
not need to be completely resolved for the elliptic fibration to
become smooth.  In other cases, the local enhancement of the gauge
group increases the rank by more than one.  By carefully analyzing the
local structure of such singularities, we can see how the resolution of
the geometry gives rise to matter in a natural generalization of the
rank one enhancement mechanism.  When the codimension one locus
in the base carrying a local factor of the gauge group itself becomes singular,
corresponding to a singular geometry for the 7-branes themselves,
matter representations are possible that cannot be realized through
elliptic fibrations whose nonabelian gauge symmetry corresponds to a
smooth component of the discriminant locus.

The specific local singularity types we consider in this paper are
motivated by global constructions.  We develop a general analysis of
F-theory Weierstrass models for theories with $SU(N)$ gauge group
localized on a  generic divisor $\sigma$ on a generic base $B$.  As $N$
increases, the set of possible
singularity structures for the Weierstrass model becomes more
complicated.  While we do not complete the general analysis of all
possibilities, we systematically show how different singularity types
can arise through different choices of algebraic structure for the
Weierstrass model.  
(This analysis complements the results of \cite{Morrison-sn}, where
the form of the Weierstrass model is determined for large $N$.)
We then apply this general analysis to the
specific cases of 6D and 4D F-theory models on bases $\P^2$ and $\P^3$.

In six dimensions, the space of allowed supergravity theories is
strongly constrained by anomalies and other simple features of the
low-energy theory \cite{Grassi-Morrison, Grassi-Morrison-2, 
universality, finite, KMT,
  tensors, 0, Seiberg-Taylor}.  We can therefore combine the classification of
theories from low-energy constraints with the analysis of singularity
structures in global F-theory models to develop a fairly complete
picture of the set of allowed matter representations in 6D quantum
supergravity theories and their realizations through F-theory.  In
particular, when $\sigma$ is a degree one curve (complex
line) on $\P^2$ we are able to reproduce
all possible matter configurations for an $SU(N)$ gauge group
compatible with anomaly conditions.  

The structure of the space of 4D
F-theory constructions and possible matter representations for an
$SU(N)$ theory is closely parallel to the 6D story; though fewer
constraints are understood from low-energy considerations in four
dimensions, similar restrictions appear on matter representations
arising in F-theory constructions.  The work presented here
represents some first steps towards a systematic understanding of the
structure of  matter in the
global space of supergravity theories arising from
F-theory compactifications.

In Section \ref{sec:local} we give the results of a local analysis for
a variety of codimension two singularities associated with matter
transforming under an $SU(N)$ gauge group.  We summarize the geometric
resolution and group theory in each case, with details of the
calculations given in an Appendix.  In Section \ref{sec:global} we
develop the general structure of  Weierstrass models with gauge
group $SU(N)$ realized on a specific divisor.  We use this general
analysis in Section \ref{sec:6D} to explicitly construct classes of
global models in 6D without tensor multiplets associated with F-theory
compactifications on $\P^2$, and in Section \ref{sec:4D} to construct
some 4D models associated with F-theory on $\P^3$.  Section
\ref{sec:conclusions} contains concluding remarks and discussion of
further directions and related open questions.

As this work was being completed we learned of related work on
codimension two singularities by Esole and Yau
\cite{Esole:2011sm}.

\section{Local analysis of codimension two singularities}
\label{sec:local}

The matter structure associated with any elliptic fibration can be
understood through a local analysis of the singularity structure of
the fibration.  Such a local analysis involves the simultaneous
resolution of all singularities in the elliptic fibration along the
lines of \cite{Katz-Morrison}.  The way in which matter arises in
F-theory can be understood from the related geometry of matter in
type IIA compactifications \cite{enhanced} and in 
M-theory compactifications as discussed by Witten \cite{Witten-mf}.
Generally, matter fields arise from $\P^1$'s in a smooth Calabi-Yau
that have been shrunk to vanishing size in the F-theory limit.  When
these $\P^1$'s arise over codimension two loci in the F-theory base
they correspond to local codimension two singularities giving rise to
localized matter.  In addition to the matter arising from local
singularities, there are also global contributions to the matter
content from $\P^1$'s that live in continuous families over the
divisor $\sigma$ in the base supporting the local factor of the gauge group.  
For example, in
a 6D model there are $g$ adjoint matter fields for $SU(N)$, where $g$
is the genus of the curve defined by $\sigma$.  We focus in this paper
on the local contributions to the matter content, though as we discuss
in Section \ref{sec:local-singular}, global matter contributions can
become local, for example when $\sigma$ develops a node.  In
this section we describe the detailed local geometry of matter in some
representations of the gauge group $SU(N)$, which is associated with a
local $A_{N -1}$ singularity on a codimension one locus (divisor)
$\sigma$ in the F-theory base $B$.

We will describe several different classes of singularities in the
discussion in this section.  We begin with the simplest types of
singularities, where the matter content can be understood through the
standard Katz--Vafa \cite{Katz-Vafa} analysis, and then consider cases
where the codimension two enhanced singularity is incompletely
resolved.  We then discuss cases where the relevant component of the
discriminant locus itself is singular.

In this paper we use explicit geometric methods to analyze F-theory
singularities.  Recently Donagi and Wijnholt \cite{Donagi-Wijnholt}
and Beasley, Heckman, and Vafa \cite{Beasley-hv}
have developed an approach to resolving singularities on intersecting
7-branes based on normal bundles and
a topological field theory on the world-volume of
the intersection.  It would be interesting to develop a better
understanding of how the analyses of this paper can be understood from
the point of view of the topological field theory framework.

To fix notation, we will be describing a local elliptic fibration
characterized by a Weierstrass model
\begin{equation}
y^2 = x^3 + fx + g \,
\label{eq:Weierstrass}
\end{equation}
where $f, g$ are local functions on a complex base $B$.  We choose
local coordinates $t, s$ on the base $B$ so that the gauge group
$SU(N)$ arises from a codimension one $A_{N -1}$ singularity on the
locus $ \sigma (t, s) = 0$.  For compactification on an elliptically
fibered Calabi-Yau threefold, $s, t$ are the only two local
coordinates needed on the base.  For 4D theories associated with a
Calabi-Yau fourfold, another coordinate $u$ is needed for the base.
This additional coordinate plays no r\^ole in the analysis in this
section.  In the simplest (smooth locus $\{\sigma = 0\}$) cases, we
can choose local coordinates with $\sigma = t$, so that the
codimension one singularity arises at $t = 0$ and the codimension two
singularity of interest arises at the coordinate $s = 0$.  In general,
the Weierstrass form (\ref{eq:Weierstrass}) of an $A_{N -1}$
singularity describes a singularity associated with a double root at
$x = x_0$ in the elliptic fiber, where $3x_0^2 + f (t = 0) = 0$, so
that it is convenient to change coordinates to $x' = x-x_0$.  The
singularity then arises at $x' = 0$, though the description of the
elliptic fibration then contains an $x^2$ term on the RHS of
(\ref{eq:Weierstrass}).  This ``Tate form'' of the description of the
elliptic fibration is often used in the mathematical analysis of
singular elliptic fibrations \cite{Tate, Bershadsky-all, Morrison-sn},
but is less convenient for a global description of the elliptic
fibration in the context of F-theory, where the Weierstrass form
allows a systematic understanding of the degrees of freedom associated
with moduli of the physical theory.
Some analyses of matter content associated with codimension two
singularities related to constructions we consider here are also
considered in \cite{Donagi:2011jy} from the spectral cover
point of view.

\subsection{Standard rank one enhancement: $A_3 \rightarrow D_4$}
\label{sec:a3}

In the simplest cases, matter arises from a codimension two
singularity in which the $A_{N -1}$ singularity, which is associated
with a rank $N -1$ gauge group, is enhanced to a singularity such as
$A_N$ or $D_N$ of one higher rank.  Such matter is characterized by
the breaking of the adjoint of the corresponding
rank $N$ group\footnote{We stress that this is not in general
an enhancement of
the gauge symmetry group.  However, the adjoint breaking provides a convenient
dictionary for the combinatorics involved, which works because in special
cases there {\em is}\/ a 
related gauge symmetry enhancement and Higgs mechanism.}
 through an embedding
of $A_{N -1}$, as described in \cite{Bershadsky-all, Katz-Vafa}.  In
particular, matter in the fundamental ({\tiny\yng(1)}) representation
can be realized through a local codimension two singularity
enhancement $A_{N -1} \rightarrow A_N$ and matter in the two-index
antisymmetric ( ${\tiny\yng(1,1)}$ ) representation (for which we will
sometimes use the shorthand notation $\Lambda^2$) can be realized
through the enhancement $A_{N -1} \rightarrow D_N$.  Matter in the
three-index antisymmetric ( ${\tiny\yng(1,1,1)}$ or $\Lambda^3$)
representation can also be realized for $SU(6), SU(7),$ and $SU(8)$
through local enhancement $A_5 \rightarrow E_6$ \cite{Bershadsky-all,
  Katz-Vafa, Grassi-Morrison-2}, $A_6\rightarrow E_7$ \cite{Katz-Vafa,
  Grassi-Morrison-2} and $A_7 \rightarrow E_8$
\cite{Grassi-Morrison-2}.

As a simple example of this kind of singularity enhancement consider a
Weierstrass model for the codimension two singularity enhancement $A_3
\rightarrow D_4$.  Though the basic physics of the matter associated
with this configuration are well understood, we go through the details
as a warmup for more complicated examples.
We consider an $A_3$ singularity on the locus $\sigma =t =
0$ with a $D_4$ singularity at $s = 0$, given by the Weierstrass form
(\ref{eq:Weierstrass}) with
\begin{eqnarray}
f & = &  -\frac{1}{3}s^4-t^2 \label{eq:a3-fg}\\
g & = &  \frac{2}{27}  s^6 +\frac{1}{3} s^2 t^2 \,.  \nonumber
\end{eqnarray}
This particular form for $f, g$ is chosen to match a form of this
singularity that appears in the general global Weierstrass analysis
in the next section of the paper.
The $A_3$ form of the singularity follows from the standard
Kodaira classification \cite{Kodaira, Morrison-Vafa-II}, 
since at generic $s \neq 0$
$f, g$
have degree $0$ in $t$, while the discriminant
\begin{equation}
\Delta = 4 f^3 + 27 g^2  = -s^4t^4-4t^6
\label{eq:discriminant}
\end{equation}
is of degree 4.
At $s = 0$, $f$ has degree 2 and the discriminant has degree 6, so we
have a $D_4$ singularity.

As mentioned above, it is convenient to change coordinates
\begin{equation}
x \rightarrow x + \frac{1}{3}s^2
\end{equation}
to move the singularity to $x = 0$.  The Weierstrass equation then
becomes
\begin{equation}
\Phi =
-y^2 + x^3 + s^2 x^2 -t^2 x = 0 \,.
\label{eq:a3}
\end{equation}
This gives a local equation for the Calabi-Yau threefold described by
an elliptic fibration in coordinates $(x, y, t, s) \in \C^2 \times \C^2$
where
$x, y$ are
(inhomogeneous) local coordinates on the elliptic fiber living in
$\P^{2, 3, 1}$ and $s, t$ are local coordinates on the base $B$.

\begin{figure}
\begin{center}
\begin{picture}(200,160)(- 100,- 90)
\put(-100,0){\makebox(0,0){\includegraphics[width=8cm]{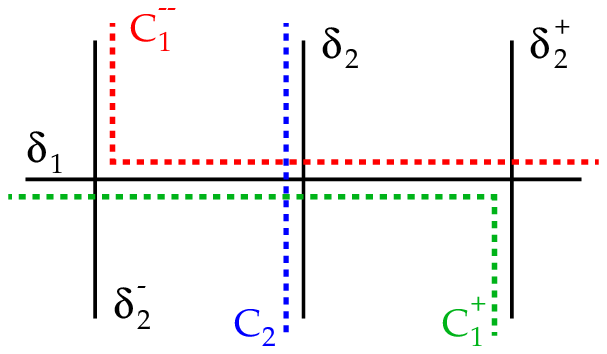}}}
\put(-100,-80){\makebox(0,0){(a)}}
\put(100,0){\makebox(0,0){\includegraphics[width=8cm]{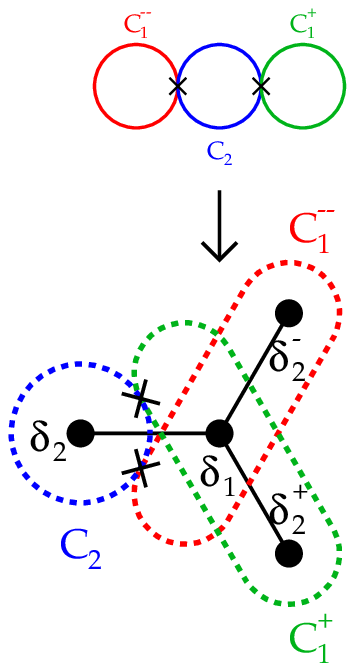}}}
\put(100,-80){\makebox(0,0){(b)}}
\end{picture}
\end{center}
\caption[x]{\footnotesize Embedding of $A_3 \rightarrow D_4$
  singularity encoded in eq.\ (\ref{eq:a3d4}).  Curves in $D_4$ are
  depicted in black solid lines, while $A_3$ curves are in colored
  dashed lines.  Two different methods are used to depict the same
  embedding.  (a) depicts each curve as a line, with intersections
  associated with crossings, as in much mathematical literature.  (b)
  depicts $D_4$ curves in Dynkin diagram notation, with nodes for
  curves and lines for intersection, and depicts $A_3$ curves as
  colored dashed curves depicting $\P^1$'s at generic $s$ and limit as
  $s \rightarrow 0$, with intersections denoted by ``x'''s.
There are four possible embeddings depending upon
  choices for codimension two resolutions.  Choice depicted has $\tau_+ = 1,
  \tau_-= 0$, according to notation in Section \ref{sec:a3-appendix} of
  Appendix, so for example $C_1^+ \rightarrow \delta_1 + \delta_2^+$
  as $s \rightarrow 0$.}
\label{f:a3d4}
\end{figure}

An explicit analysis of the singularity resolution of the Calabi-Yau
threefold defined by \eq{eq:a3} is given in Section
\ref{sec:a3-appendix} of the Appendix.  Even in this rather simple
case, the details of the resolution are slightly intricate.  At a
generic point $s \neq 0$ along the $A_{3}$ singularity $\sigma$, a
blow-up in the transverse space gives two $\P^1$'s ($C_{\pm}$) fibered
over $\sigma$, which intersect at a singular point for each $s$.  A
further blow-up gives a third $\P^1$ ($C_2$) fibered over $\sigma$,
which intersects each of $C_\pm$, giving a realization of the Dynkin
diagram $A_3$ in terms of the intersections of these curves.  At $s =
0$, the resolution looks rather different.  The first blow-up gives a
single curve $\delta_1$, which both $C_+, C_-$ approach in the limit
$s \rightarrow 0$.  A further blow-up at a singular point on
$\delta_1$ gives $\delta_2 \sim C_2$, and codimension two
conifold-type double point singularities occur at two other points on
$\delta_1$.  Each of these codimension two singularities has two
possible resolutions, giving four possible smooth Calabi-Yau threefold
structures related by flops.  In each  resolution an
additional $\P^1$ is added at $s = 0$, completing the $D_4$ Dynkin
diagram.  An example of how the curves $C_a$ at generic $s$ converge
to the curves $\delta_b$ at $s = 0$ for one of the four combinations
of  resolutions is shown graphically in Figure~\ref{f:a3d4}.

The additional matter associated with the $D_4$ can be understood by
embedding $A_3 \subset D_4$ and decomposing the adjoint of $D_4$ into
irreducible representations of $A_3$.  The roots in the adjoint of
$D_4$ correspond to distinct $\P^1$'s at $s = 0$ in the resolved
Calabi-Yau.  The subset of these roots corresponding to the adjoint of
$A_3$ are associated with the $SU(4)$ vector bosons, and the remainder
are matter fields.  The adjoint of $D_4$ decomposes into irreducible
representations of $A_3$ as ${\bf 28} \rightarrow {\bf 15} + {\bf 6} +
{\bf \bar{6}} + {\bf 1}$.  In a 6D theory, matter hypermultiplets live
in quaternionic representations of the gauge group.  The ${\bf 6}$ and
${\bf \bar{6}}$ combine into a single quaternionic matter
hypermultiplet in the $\Lambda^2$ representation of $A_3$ in 6D.  An
easy way to see that this representation appears is from the Dynkin
diagram description of the embedding $A_3 \subset D_4$.
(The embedding shown in Figure~\ref{f:a3d4} is equivalent to this
embedding under an isomorphism of $D_4$.)
\begin{center}
\begin{picture}(200,50)(- 100,- 35)
\put(0,0){\makebox(0,0){$\longrightarrow$}}
\put(-60,-5){\line( 1, 0){20}}
\put(40,-5){\line( 1, 0){20}}
\multiput(- 60,-5)(10,0){3}{\circle*{4}}
\multiput(40,-5)(10,0){3}{\circle*{4}}
\put(50,5){\circle{4}}
\multiput(50,-5)(0,4){3}{\line(0,1){2}}
\put(-50,-20){\makebox(0,0){$A_3$}}
\put(50,-20){\makebox(0,0){$D_4$}}
\end{picture}
\end{center}
The Dynkin weight $[0, 1, 0]$ is the highest weight of the $\Lambda^2$
representation of $A_3$.  The $\P^1$ associated with this state is
precisely the extra (empty) node added to form $D_4$ from $A_3$ in
this embedding.  The weight of this state can be determined from the
intersection numbers of this $\P^1$ with the roots of $A_3$; the
additional $\P^1$ has intersection number 1 with the middle root of
$A_3$ and no intersection with the other roots. (See \cite{Slansky}
for a review of the notation of Dynkin weights and the relevant group
theory.)

\subsection{Incomplete and complete resolutions}
\label{sec:a5}

We now consider a slightly more complicated set of enhancements of an
$A_{N -1}$ codimension one singularity.  In this case we consider the
enhancement of $A_5$ by various types of local singularities and the
associated matter content.  We will begin with the example of $A_5$
enhanced to $D_6$ through a standard rank one enhancement quite
similar to the preceding analysis of $A_3 \rightarrow D_4$.  This
again gives a matter field in the $\Lambda^2$ antisymmetric
representation.  We will then consider the effect of a local $E_6$
singularity.  Depending upon the degree of vanishing of certain terms
in the local defining equation, the $E_6$ can either be incompletely
resolved or can be completely resolved in the threefold.  The $E_6$
singularity gives rise to matter in the three-index antisymmetric
($\Lambda^3$) representation; in the 6D context we get a half or
full hypermultiplet in this representation depending on whether the
singularity is completely resolved.

In each case, we choose a non-generic Weierstrass model, with a
specific form motivated by the global analysis carried out in the
following section.

\subsubsection{Enhancement $A_5 \subset D_6$}

We begin with the Weierstrass coefficients
\begin{eqnarray}
f & = &  -\frac{1}{3}s^4 -2s^3t + (2s^2 -3)t^2 + 3 t^3,
 \label{eq:a5-d6-fg}\\
g & = &  \frac{2}{27}  s^6 +
\frac{2}{3}  s^4 t
+ (2 s^2-\frac{2}{3}  s^4) t^2 +(2-3 s^2) t^3 + (s^2 -3)t^4 \,.  \nonumber
\end{eqnarray}
These describe an $A_5$ singularity on the locus $t = 0$ enhanced to a
$D_6$ singularity at $s = 0$, where the orders of vanishing of $f, g,
\Delta$ are 2, 3, 6.  Changing variables through
\begin{equation}
x \rightarrow x + \frac{1}{3}s^2+ t
\end{equation}
gives the local equation
\begin{equation}
\Phi = -y^2 + x^3 +s^2 x^2 +3x^2 t +
 3t^3x + 2s^2 t^2 x + s^2 t^4  = 0\,.
\label{eq:phi-a5-d6}
\end{equation}
\vspace*{0.05in} 
An analysis much like that of $A_3 \subset D_4$,
summarized in Section \ref{sec:a5-d6} of the Appendix, shows that this
singularity is resolved to give a set of curves with $D_6$ structure
at $s = 0$, giving matter in the $\Lambda^2$ representation of $A_5$ with
highest weight vector having Dynkin indices $[0, 1, 0, 0, 0]$.

\subsubsection{Enhancement $A_5 \subset E_6$}

\begin{figure}
\begin{center}
\begin{picture}(200,160)(- 100,- 90)
\put(-100,10){\makebox(0,0){\includegraphics[width=8cm]{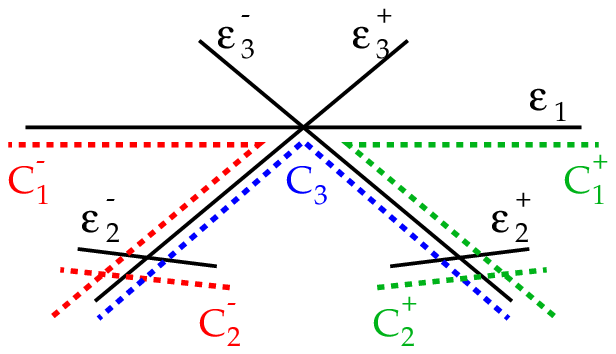}}}
\put(-100,-90){\makebox(0,0){(a)}}
\put(100,0){\makebox(0,0){\includegraphics[width=7cm]{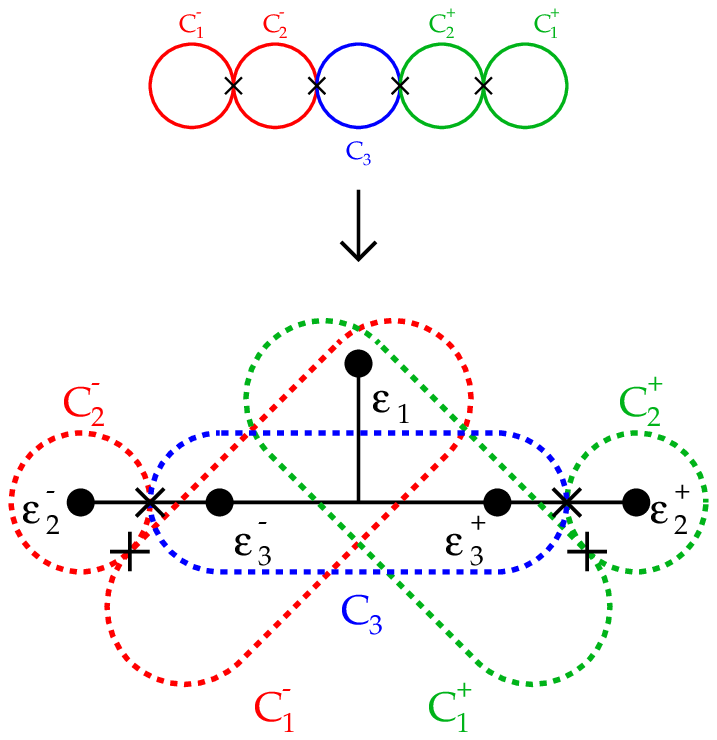}}}
\put(100,-90){\makebox(0,0){(b)}}
\end{picture}
\end{center}
\caption[x]{\footnotesize  Embedding $A_5 \rightarrow E_6$ with
  incomplete resolution of $E_6$ singularity in threefold.}
\label{f:a5-e6x}
\end{figure}

\begin{figure}
\begin{center}
\begin{picture}(200,180)(- 100,- 90)
\put(-100,10){\makebox(0,0){\includegraphics[width=8cm]{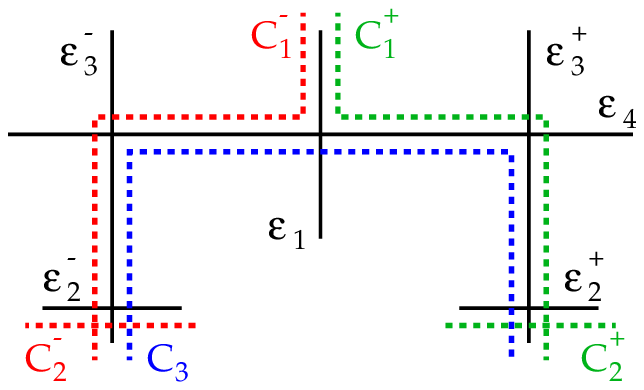}}}
\put(-100,-90){\makebox(0,0){(a)}}
\put(100,0){\makebox(0,0){\includegraphics[width=7cm]{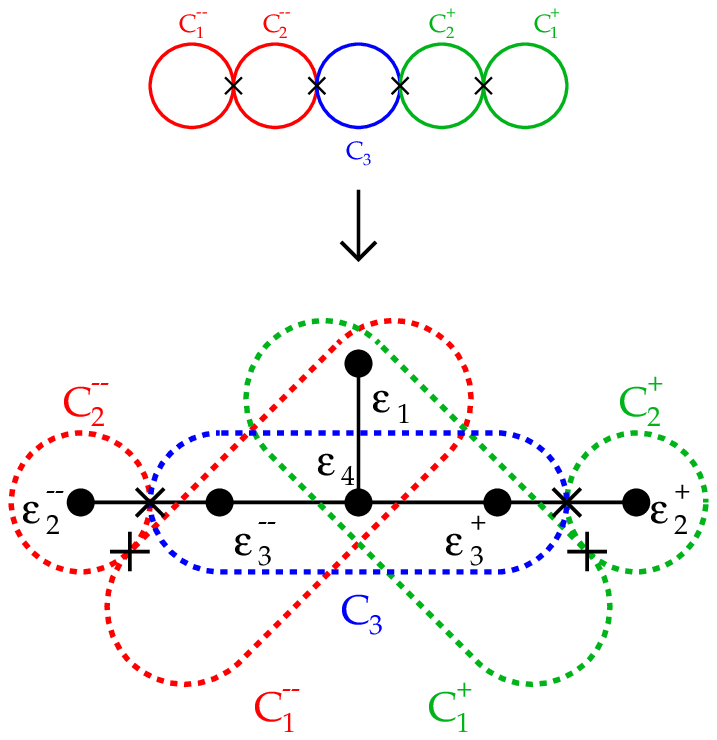}}}
\put(100,-90){\makebox(0,0){(b)}}
\end{picture}
\end{center}
\caption[x]{\footnotesize  Embedding $A_5 \rightarrow E_6$ with
complete resolution of $E_6$ singularity in threefold.}
\label{f:a5-e6}
\end{figure}

We now consider a  situation where $A_5$ is
enhanced to $E_6$.  The local model we consider is closely related to
(\ref{eq:phi-a5-d6}).
We begin with the Weierstrass coefficients
\begin{eqnarray}
f & = &  -\frac{1}{3}\betah^4 -2 \betah^3t + (2\betah -3 \betah^2)t^2 + 3 t^3,
 \label{eq:a5-e6-fg}\\
g & = &  \frac{2}{27}  \betah^6 +
\frac{2}{3}  \betah^5 t
+ (2 \betah^4-\frac{2}{3}  \betah^3) t^2 +(2 \betah^3-3 \betah^2) t^3 + (1
-3 \betah)t^4 \,.  \nonumber
\end{eqnarray}
Changing variables through
\begin{equation}
x \rightarrow x + \frac{1}{3}\betah^2+  \betah t
\end{equation}
gives
\begin{equation}
\Phi = -y^2 + x^3 + \betah^2 x^2 +3 \betah x^2 t +
 3t^3x + 2 \betah t^2 x + t^4  = 0\,.
\label{eq:phi-a5-e6}
\end{equation}
We describe explicit global 6D models in which this singularity
structure arises in Section \ref{sec:6D}.  In (\ref{eq:phi-a5-e6}),
the parameter $\betah$ can be either $\betah = s$ or $\betah = s^2$.  The
detailed analysis of the singularity resolution in both cases is
carried out in Section \ref{sec:a5-e6} of the Appendix.  To understand
the results of this analysis it is helpful to clarify the structure of
the $E_6$ singularity at $s = 0$.  The Kodaira classification of
singularities is really only applicable in the context of codimension
one singularities.  For generic $s$, we can take a slice at constant
$s$, giving a codimension one singularity of type $A_5$ on each slice
intersecting the curve at $t = 0$.  To determine the type of
singularity at $s = t = x =y =0,$ we are considering a slice at $s =
0$.  Just because there is a singularity in this slice, however, does
not mean that the full Calabi-Yau threefold is singular.  In
particular, in the case at hand, when $\betah = s$, systematically
blowing up the singularity at the origin allows the Calabi-Yau
threefold to be smoothed before the full $E_6$ singularity has been
resolved.  At the final stage of this resolution process, there is an
apparent singularity in the slice at $s = 0$ but the full threefold
has no singularity.  A diagram depicting the blown-up $\P^1$'s away
from $s = 0$ ($C_a$'s) and at $s = 0$ ($\epsilon_b$'s) for the
incomplete $E_6$ resolution from $\betah  = s$
is shown in
Figure~\ref{f:a5-e6x}.
When $\betah = s^2$, the full $E_6$ singularity is resolved, giving the
configuration depicted in Figure~\ref{f:a5-e6}.
Although this explicit singularity resolution gives an embedding of
$A_5 \subset E_6$ with a somewhat unconventional appearance, this
embedding is unique up to automorphisms of $E_6$, so is equivalent to
the embedding associated with extending the Dynkin diagram $A_5$ by
adding a new node attached to middle node of the $A_5$ to form the
$E_6$ diagram.
\begin{center}
\begin{picture}(200,70)(- 100,- 35)
\put(0,0){\makebox(0,0){$\longrightarrow$}}
\put(-70,-5){\line( 1, 0){40}}
\put(30,-5){\line( 1, 0){40}}
\multiput(- 70,-5)(10,0){5}{\circle*{4}}
\multiput(30,-5)(10,0){5}{\circle*{4}}
\put(50,5){\circle{4}}
\multiput(50,-5)(0,4){3}{\line(0,1){2}}
\put(-50,-20){\makebox(0,0){$A_5$}}
\put(50,-20){\makebox(0,0){$E_6$}}
\end{picture}
\end{center}

Let us now consider the matter content in each of these situations.
In the fully resolved $E_6$, we have the usual story of rank one
enhancement and the adjoint of $E_6$ decomposes under
$A_5 \subset E_6$
as
\begin{equation}
{\bf 78} = 3 \cdot {\bf 1} + {\bf 35} + 2 \cdot {\bf 20} \,.
\end{equation}
This gives matter in the 3-index antisymmetric ($\Lambda^3$) {\bf 20}
representation of $A_5$.  Again, the appearance of this matter
representation is apparent from the Dynkin index $[0, 0, 1, 0, 0]$
associated with the intersection of the added node with the original
nodes of the $A_5$.  The possibility of this kind of matter associated
with a local $E_6$ enhancement was previously discussed in
\cite{Bershadsky-all, Katz-Vafa, Grassi-Morrison-2}.  In a 6D theory,
as in the $\Lambda^2$ matter story, the two {\bf 20}'s combine into a single
full hypermultiplet.  Because the ${\bf 20}$ is by itself already a
quaternionic representation of $A_5$, however, this can also be
thought of as two half-hypermultiplets.

\begin{figure}
\begin{center}
\begin{picture}(200,80)(- 90,- 40)
\multiput(-24,15)(8,0){7}{\circle*{4}}
\multiput(-40, 0)(8,0){11}{\circle*{4}}
\multiput(-24,-15)(8,0){7}{\circle*{4}}
\put(60, 15){\makebox(0,0){{\bf 20}}}
\put(60, 0){\makebox(0,0){{\bf 35}}}
\put(60, -15){\makebox(0,0){{\bf 20}}}
\put(70,15){\vector(1,-1){10}}
\put(70,0){\vector(1,0){10}}
\put(70,-15){\vector(1,1){10}}
\put(110,0){\makebox(0,0){ {\bf 35} + {\bf 20}}}
\put(-50,-15){\vector(0,1){30}}
\put(-65,-10){\makebox(0,0){$\epsilon_4$}}
\end{picture}
\end{center}
\caption[x]{\footnotesize A schematic depiction of the decomposition
  of the adjoint of $E_6$ under the action of $A_5$.
The action of $A_5$ is taken to be in the horizontal direction.  The
root $\epsilon_4$ is perpendicular to all roots of $A_5$.  In the
incompletely resolved $E_6$, a projection is taken in the $\epsilon_4$
direction that combines the two {\bf 20}'s of $A_5$ into a single
half hypermultiplet.}
\label{f:collapse}
\end{figure}
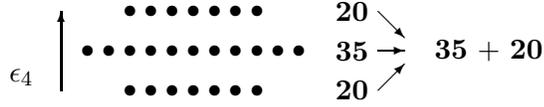

Now we consider the case where the $E_6$ is incompletely resolved.  In
this case, the set of roots of $E_6$ are not all associated with
$\P^1$'s in the full Calabi-Yau over the point $s = 0$.  Thus, the
amount of matter is reduced.  The root of $E_6$ that is not blown up,
associated with the curve $\epsilon_4$ in the complete resolution
depicted in Figure~\ref{f:a5-e6}, is orthogonal to all roots of $A_5$.
For example, $\epsilon_4 \cdot C_1^+ = \epsilon_4 \cdot ( \epsilon_1 +
\epsilon_3^+ + \epsilon_4) = -2 + 1 + 1 = 0$.  We can therefore
describe the matter content of the incompletely resolved $E_6$ by
projecting in the direction parallel to $\epsilon_4$.  This
collapses the two ${\bf 20}$'s into a single matter representation
(see Figure~\ref{f:collapse}).
In the 6D theory this gives a half-hypermultiplet in the $\Lambda^3$
representation.  It was also noted in \cite{Katz-Vafa} that the
appearance of a quadratic parameter like $\betah = s^2$ in the defining
equation of the singularity is associated with a pair of
half-hypermultiplets in certain situations; this observation matches
well with the appearance of a single half-hypermultiplet when $s^2$ is
replaced with $s$.

It is interesting to understand how the intersection properties of the
$C$ curves from $A_5$ are realized in the incomplete $E_6$ resolution.
The detailed expansion of the $C$'s in terms of the roots $\epsilon$
of the $E_6$ is given in \eq{eq:c-e}, and shown graphically in
Figure~\ref{f:a5-e6x}.  In the incompletely
resolved $E_6$, we can consider the geometry of the slice at $s = 0$
containing blown-up $\P^1$'s associated with all roots of $E_6$ other
than $\epsilon_4$.  In this slice, there is a $\Z_2$ singularity at
the intersection point of $\epsilon_1, \epsilon_3^\pm$.  This point
contributes only 1/2 to the Euler characteristic of spaces in which it is
contained.  Each of the curves intersecting the point consequently has
a self-intersection given by $\epsilon_1 \cdot \epsilon_1 = -3/2$,
{\it etc.} and the intersection between each pair of curves meeting at this point is 1/2, so
$\epsilon_1 \cdot \epsilon_3^\pm = 1/2$, {\it etc.}.  We see that the
linear combinations of these singular curves spanned by the $C$'s
preserve the correct intersection rules for $A_5$; for example,
\begin{equation}
C_1^+ \cdot C_3 = (\epsilon_1 + \epsilon_3^+) \cdot (\epsilon_3^+ +
\epsilon_3^-) = -3/2 + 3 (1/2) = 0,
\end{equation}
\begin{equation}
C_3 \cdot C_3 = (\epsilon_3^+ +
\epsilon_3^-)\cdot (\epsilon_3^+ +
\epsilon_3^-)= 2 (-3/2) + 2 (1/2) = -2 \,.
\end{equation}

We expect that there are many types of codimension two singularities
that can appear in F-theory with analogous descriptions in terms of
incomplete resolutions.  In Section
\ref{sec:6D} we describe global 6D F-theory models in which this kind
of incomplete resolution appears explicitly, affecting the matter
content of the theory.

\subsection{Matter on a singular 7-brane}
\label{sec:local-singular}

For the fundamental and multi-index antisymmetric representations of
$SU(N)$ that we have studied so far, the associated F-theory geometry
involves the enhancement at a codimension two locus of an $A_{N -1}$
singularity living on a 7-brane that itself is wrapped on a smooth
codimension one locus in the base.  Other kinds of representations can
arise when the 7-branes are wrapped on a singular divisor.  In
\cite{Sadov}, Sadov gave some evidence suggesting that a two-index
symmetric ( ${\tiny\yng(2)}$ or Sym${}^2$) representation of $SU(N)$
should arise when the gauge group is realized on a codimension one
space having an  ordinary double point singularity.  The
connection between matter representations and geometric singularities
can be made much more general from analysis of anomaly cancellation in
6D theories.  We describe the general connection that we expect
between matter representations and singularities in the 7-brane
configuration, and then describe in some detail the case of the
ordinary double point singularity from this point of view.

\subsubsection{Representation theory and singularities}

It was found in \cite{0} that associated with each representation of
$SU(N)$ there is a numerical factor $g_R$ that  corresponds in a
6D F-theory model to a contribution to the genus of the divisor
associated with the $SU(N)$ local factor.  The analysis in \cite{0} was based on
compactifications on $\P^2$, but the result can be stated more
generally.  From the anomaly cancellation conditions for an F-theory
construction on an arbitrary base (see \cite{WT-TASI} for a review),
the genus $g$ of the curve $C$ on which the 7-branes associated with
any $SU(N)$ local factor of the gauge group 
are wrapped can be written in terms of
a sum over contributions from each matter representation
\begin{equation}
2g-2 = (K + C) \cdot C= \sum_{R}x_R g_R -2 \,,
 \label{eq:Euler}
\end{equation}
where $x_R$ is the number of matter hypermultiplets in representation
$R$, and  the genus contribution of a given
representation is defined to be
\begin{equation}
g_R = \frac{1}{12}\left(2 C_R + B_R -A_R \right) \,.
\label{eq:genus}
\end{equation}
In this formula, $A_R, B_R, C_R$ are group theory coefficients defined
through
\begin{align}
\tr_R F^2 & = A_R  \tr F^2 \\
\tr_R F^4 & = B_R \tr F^4+C_R (\tr F^2)^2 \,,
\end{align}
where $\tr_R$ denotes the trace in representation $R$, while $\tr$
without a subscript denotes the trace in the fundamental
representation.  A table of group theory coefficients and genera for
some simple $SU(N)$ representations appears in \cite{0}; we reproduce
here the part of the table describing representations with nonzero
genus in Table~\ref{t:coefficients}.  All single-column antisymmetric
representations ($\Lambda^2$, $\Lambda^3$, \ldots) have vanishing
genus.
\begin{table}
\centering
\begin{tabular}{|c|c|c|c|c|c|c|}
\hline
 Rep. & Dimension & $A_R$ & $B_R$ & $C_R$  & $g_R$\\
\hline
 Adjoint & $N^2-1$ & $2N$ & $2N$ & 6 & 1\\
 ${\tiny\yng(2)}$ & $ \frac{N(N+1)}{2} $ & $ N+2 $ & $ N+8 $ & 3 & 1\\
 ${\tiny\yng(2,1)}$ & $ \frac{N(N^2-1)}{3} $& $ N^2-3 $& $ N^2-27 $& $
 6N $ &  $N -2 $\\
 ${\tiny\yng(3)}$ & $ \frac{N(N+1)(N+2)}{6} $ &$ \frac{N^2+5N+6}{2}
 $&$ \frac{N^2+17N+54}{2} $ & $ 3N+12 $ &  $N + 4$\\
 ${\tiny\yng(2,2)}$ &  $ \frac{N^2(N+1)(N-1)}{12} $ &
 $\frac{N(N-2)(N+2)}{3} $ & $\frac{N(N^2-58)}{3}$ & $3(N^2+2)$  &  $\frac{(N-1)(N-2)}{2}$ \\[0.07in]
\hline
\end{tabular}
\caption{Values of the group-theoretic coefficients $A_R, B_R, C_R$,
  dimension and genus
for some representations of $SU(N)$, $N \geq 4$.}
\label{t:coefficients}
\end{table}

While we have described so far the relationship between representation theory
and geometry of singularities  only for $SU(N)$ local factors and
representations, a similar result holds for any simple local factor
of the gauge group
with the inclusion of appropriate numerical factors depending on the
normalization of the trace.
For a general gauge group the genus contribution is
\begin{equation}
g_R = \frac{1}{12}\left(2 \lambda^2 C_R + \lambda B_R -\lambda A_R \right) \,,
\label{eq:genus-general}
\end{equation}
where $\lambda$ is a group-dependent normalization factor, with
$\lambda_{SU(N)} = 1, \lambda_{SO(N)} = 2$, etc.; values of $\lambda$
for all simple groups are listed in \cite{tensors}.

We now review some elementary features of plane curves (complex curves
in $\P^2$) that clarify the connection of the group theory structure
just described with singularities in F-theory.  For more background in
the basic algebraic geometry of plane curves see, e.g., \cite{Perrin}.  In
algebraic geometry, a smooth plane curve is characterized by two
invariants: the degree $b$ of the polynomial defining the
curve in $\P^2$, and the
genus $g$ of the curve, which is related to the Euler characteristic of the
curve through the usual relation
\begin{equation}
\chi = 2-2g
\end{equation}
For a smooth plane curve, the degree and
genus are related by
\begin{equation}
2g =  (b -1) (b-2).
\label{eq:arithmetic-genus}
\end{equation}
Thus, lines ($b = 1$) and conics ($b = 2$) have genus 0, smooth cubics
($b = 3$) are elliptic curves of genus 1, curves of degree 4
have genus 3, etc.

Using inhomogeneous coordinates $t, s$ on $\P^2$, a curve $f (t, s) =
0$ is singular at any point where
\begin{equation}
\partial f/\partial t = \partial f/\partial s = 0.
\end{equation}
For example, the cubic 
\begin{equation}
f (t, s) = t^3 + s^3-st = 0
\label{eq:singular-cubic}
\end{equation}
is singular at the point $(t, s) = (0, 0)$, and locally takes the form
$st = 0$, describing two lines crossing at a point.  

For a singular curve, there are two distinct notions of genus that
become relevant.  The {\it arithmetic} genus is given by
(\ref{eq:arithmetic-genus}) for any curve, singular or nonsingular.
The {\it geometric} genus (which we denote by $p_g$) is the topological
genus of a curve after all singularities have been appropriately
smoothed.  For example, the singularity in (\ref{eq:singular-cubic})
is known as an {\it ordinary double point} singularity, where two
smooth branches of the curve cross at a point.  This singularity can
be removed by blowing up the origin to a $\P^1$, which separates the
two points, giving a curve of geometric genus 0.  In general, the
arithmetic and geometric genera of a plane curve $C$ with multiple
singularities are related through
\begin{equation}
g =  (b-1) (b-2)/2 = p_g + \sum_P \frac{m_P(m_P-1)}{2},
\label{eq:genus-relation}
\end{equation}
where the sum is over\footnote{There is an important subtlety here:
after blowing up a singular point, there may still be singular points
in the inverse image of the original point, and they must be blown
up as well, {\em ad infinitum.}  The sum in (\ref{eq:genus-relation})
must include these ``infinitely near'' points.}
 all singular points $P$ in $C$, and $m_P$ is the
multiplicity of the singularity at $P$.  The multiplicity of an
ordinary singularity where $k$ branches of the curve cross at a common
point is $k$.  It is easy to see that deforming such a singularity
leads to $k (k -1)/2$ ordinary double point singularities, each of
which contributes one to the genus.  More generally, the multiplicity
of a singularity in a plane curve is given by the lowest power of a
monomial appearing in the polynomial defining the curve in local
coordinates around the singularity.  For example, for degree 3 curves,
in addition to the ordinary double point type of singularity
encountered in (\ref{eq:singular-cubic}), a {\it cusp} (non-ordinary)
double point singularity can arise at points like the origin in the
cubic
\begin{equation}
f (t, s) = t^3-s^2 = 0 \,.
\label{eq:cubic-cusp}
\end{equation}
The multiplicity of such a cusp
singularity is 2; this cusp can be found as a degenerate limit of the
class of cubics with ordinary double point singularities $t^3 + at^2
-s^2 = 0$ as $a \rightarrow 0$.  Note that
a curve of geometric genus 0 is a {\it rational} curve, meaning that
the curve can be parameterized using rational functions.  For example,
(\ref{eq:cubic-cusp}) has arithmetic genus 1 but geometric genus 0,
and can be parameterized as $t = a^2, s = a^3$.
For higher degree curves, more exotic
types of singularities can arise with higher intrinsic multiplicities.
While an ordinary double point singularity is resolved by a single blow-up,
as the singularity becomes more extreme, the point must be blown up
more times to completely resolve the singularity.
From (\ref{eq:genus-relation}), we see that the total arithmetic genus
of a curve has a contribution from the geometric genus and also a
contribution from the various singular points in the curve.  

We now return to the discussion of matter and singularities in the
F-theory context.  The genus appearing in \eq{eq:Euler} is the
arithmetic genus.  For a local factor of the gauge group 
associated with 7-branes
on a smooth curve, there are $g$ matter fields in the adjoint
representation, associated with $\P^1$'s in the resolved space that
are free to move over the curve of genus $g$.  Since the adjoint
representation has $g_R = 1$, these non-localized adjoint matter
fields saturate \eq{eq:genus-relation}; a gauge group realized on a
smooth curve can thus only have local matter in the fundamental and
multi-index antisymmetric representations.  If the gauge group lives
on a singular curve, however, the number of adjoint representations is
given by $p_g$, with the type of singularities in the curve
determining the types of additional matter that can arise.  From
(\ref{eq:genus-relation}), we expect that a matter representation $R$
will be associated with a localized singularity in $\sigma$
contributing $g_R$ to the genus.  This gives a clear picture of how
matter representations should be associated with singular divisor
classes in F-theory.

For example, consider the symmetric (Sym${}^2$) representation of $SU(N)$.
This representation has $g_{{\rm Sym}{}^2} = 1$ and should be associated with a
singularity of multiplicity 1.  This matches with Sadov's prediction
that such matter should be associated with an ordinary double point 
 in $\sigma$.  We analyze this type of singularity in detail in
the following section.  As another example, consider the ``box''
( ${\tiny\yng(2,2)}$ ) representation of $SU(4)$.  From
Table~\ref{t:coefficients}, we see that this representation has genus
$g_R = 3$.  
Thus, we expect that it will be produced by a singularity of
multiplicity 3 in the divisor locus carrying a stack of 7-branes in a
6D F-theory model on $\P^2$.
In \cite{0} it was shown that this representation can
arise in apparently consistent 6D supergravity models with $SU(4)$ gauge group 
and no tensor multiplets.  We discuss this representation further in
Section \ref{sec:box}.

Although the preceding discussion was based on the analysis of 6D
supergravity models, where anomaly cancellation strongly constrains the
range of possible models, the connection between the group theoretic genus
contribution of a given matter representation and the corresponding
singularity type in the F-theory picture should be independent of
dimension.  Thus, in particular, the same correspondence will relate
matter in 4D supergravity models to localized codimension two
singularity structures in F-theory compactifications on a Calabi-Yau
threefold, just as the Kodaira classification describes
gauge groups based on codimension one singularities in all dimensions.

\subsubsection{Ordinary double point singularities}
\label{sec:double}

We now consider the simplest situation where the locus $\sigma$ on
which the 7-branes are wrapped itself becomes singular in a 6D
F-theory model.  This occurs when $\sigma$ contains an ordinary double
point singularity, such as arises at the origin for the curve $u^3 +
u^2 -v^2 = u^3 + (u + v) (u-v) = 0$.  Locally, an ordinary double
point singularity takes the form $st = 0$ in a local coordinate
system; here this is the case with $s = u + v, t = u-v$.  The geometry
and physics of such an intersection is well-known.  A stack of
7-branes associated with an $A_{N -1}$ codimension one singularity
that has a transverse intersection with a stack of 7-branes
associated with an $A_{M -1}$ singularity gives rise to matter in a
bifundamental representation
\begin{equation}
(N, \bar{M}) + (\bar{N}, M) \; \;
{\rm or} \; \;
(N, M) + (\bar{N}, \bar{M}).
\label{eq:bifundamental}
\end{equation}
For two branes that intersect each other in only one place, and do not
intersect other branes, these two representations are effectively
indistinguishable, being equivalent under a redefinition of
the gauge group on one of the branes.  When the branes have multiple
intersections, or are identified, however, the relative structure of
representations from one intersection to another (or from one branch
of the brane to another) means that these two distinct types of
bifundamental representations must be distinguished.  When the two
stacks of 7-branes are actually the same, corresponding to a
self-intersection of $\sigma$, the resulting representation of $SU(N)$
is either an adjoint hypermultiplet, or a symmetric and an
antisymmetric hypermultiplet ($\Lambda^2$ + Sym${}^2$)
\begin{eqnarray}
1 + {\rm adj}: &  & {\rm singlet}  (1) + {\rm adjoint} (N^2 -1)  \nonumber\\
 & {\rm or} &  \label{eq:crossing-options}\\
\Lambda^2 + {\rm Sym}^2: &  & {\tiny\yng(1,1)} \;(N (N -1)/2)+
{\tiny\yng(2)}\;(N (N +1)/2)\,.
\nonumber
\end{eqnarray}
To understand which of these representations is realized, and to
connect with the general discussion of matter and singularities, it is
helpful to go through the F-theory singularity analysis in a similar
fashion to that done for the singularities analyzed above.  For
concreteness, we describe the self-intersection of a curve $\sigma$
carrying an $A_3$ singularity in a 6D model.

To describe an $SU(4)$ gauge group on a singular divisor class
$\sigma$, we can substitute $s\rightarrow 1, t \rightarrow \sigma$ in
the equation (\ref{eq:a3}) for an $A_3$ singularity
\begin{equation}
\Phi =
-y^2 + x^3 + x^2 -\sigma^2 x = 0 \,.
\end{equation}
For the local ordinary double point $\sigma = st$, we have
\begin{equation}
\Phi =
-y^2 + x^3 + x^2 -s^2 t^2 x = 0 \,.
 \label{eq:ordinary-double}
\end{equation}
This defines a Calabi-Yau threefold that is singular along the lines
$s = 0$ and $t = 0$ with an enhancement to $A_7$ at the point $s = t =
0$.  We can resolve the singularity systematically by blowing up the
curves along $t = 0$, $s = 0$ and at the origin.  The details are
described in Section \ref{sec:ordinary-double-appendix}.  The result of this
analysis is that the 3 $\P^1$'s giving the $A_3$ structure along each
of the curves $s = 0, t = 0$ are embedded into two orthogonal $A_3$
subgroups of the $A_7$ Dynkin diagram.  The embedding found from
explicit singularity resolution is equivalent to the canonical
embedding of $SU(4) \times SU(4) \subset SU(8)$ depicted in terms of
Dynkin diagrams as
\begin{center}
\begin{picture}(200,50)(- 100,- 35)
\put(0,0){\makebox(0,0){$\longrightarrow$}}
\put(-100,0){\line( 1, 0){20}}
\put(-40,0){\line( 1, 0){20}}
\put(30,0){\line( 1, 0){60}}
\multiput(- 100,0)(10,0){3}{\circle*{4}}
\multiput(- 40,0)(10,0){3}{\circle*{4}}
\multiput(30,0)(10,0){3}{\circle*{4}}
\multiput(70,0)(10,0){3}{\circle*{4}}
\put(60,0){\circle{4}}
\put(-60,0){\makebox(0,0){$\times$}}
\put(-60,-20){\makebox(0,0){$A_3\times A_3$}}
\put(60,-20){\makebox(0,0){$A_7$}}
\end{picture}
\end{center}
We can then decompose the adjoint of $A_7$
as usual to get the matter content.
If the two $SU(4)$ gauge groups were
independent, this would give   one of the bifundamental
representations \eq{eq:bifundamental}.  When the two $A_3$ singularity
loci are connected, however, which of the matter representations
\eq{eq:crossing-options} are realized depends upon the geometry of
$\sigma$.  Locally, the 3 $\P^1$'s associated with simple roots of
$A_3$ on one branch can be labeled with 1, 2, 3.  When this labeling
is followed around $\sigma$ onto the second branch, we have an
embedding of a single $A_3$ through
\begin{equation}
A_3 \rightarrow
A_3 \times A_3 \rightarrow A_7
\end{equation}
that can be realized through either of the
two possibilities
\begin{center}
\begin{picture}(200,70)(- 30,- 35)
\put(-10,0){\makebox(0,0){$\longrightarrow$}}
\put(-70,0){\line( 1, 0){20}}
\put(30,0){\line( 1, 0){60}}
\put(140,0){\line( 1, 0){60}}
\multiput(- 70,0)(10,0){3}{\circle*{4}}
\multiput(30,0)(10,0){3}{\circle*{4}}
\multiput(70,0)(10,0){3}{\circle*{4}}
\multiput(140,0)(10,0){3}{\circle*{4}}
\multiput(180,0)(10,0){3}{\circle*{4}}
\put(60,0){\circle{4}}
\put(170,0){\circle{4}}
\put(-60,-20){\makebox(0,0){$A_3$}}
\put(60,-20){\makebox(0,0){$A_7$}}
\put(170,-20){\makebox(0,0){$A_7$}}
\put(-70,10){\makebox(0,0){1}}
\put(-60,10){\makebox(0,0){2}}
\put(-50,10){\makebox(0,0){3}}
\put(30,10){\makebox(0,0){1}}
\put(40,10){\makebox(0,0){2}}
\put(50,10){\makebox(0,0){3}}
\put(70,10){\makebox(0,0){1}}
\put(80,10){\makebox(0,0){2}}
\put(90,10){\makebox(0,0){3}}
\put(140,10){\makebox(0,0){1}}
\put(150,10){\makebox(0,0){2}}
\put(160,10){\makebox(0,0){3}}
\put(200,10){\makebox(0,0){1}}
\put(190,10){\makebox(0,0){2}}
\put(180,10){\makebox(0,0){3}}
\put(115,0){\makebox(0,0){or}}
\end{picture}
\end{center}
These two possibilities correspond to the two matter options
\eq{eq:crossing-options}.

Thus, we see that an ordinary double point singularity in $\sigma$ can
either be associated with an adjoint plus a singlet, or a symmetric
and an antisymmetric matter multiplet.  In each case, the contribution
through \eq{eq:genus} to the genus is 1, so either possibility is
consistent with the general picture of the association between
geometry and group theory.

Which of the possible representations is realized, however, is
determined by nonlocal features of the geometry.
To see which of the embeddings from the diagram above is
realized it is necessary to track the labeling of the $A_3$ roots
around a closed path in $\sigma$ connecting the two branches that
intersect.  The information in the orientation of the ordering of
these roots amounts to an additional $\Z_2$ of information contained in
the structure of any brane.  It is interesting to note that this
degree of freedom is present in any configuration of type II D-branes,
although it is not generally discussed.

\begin{figure}
\begin{center}
\begin{picture}(200,150)(- 100,- 70)
\put(0,0){\makebox(0,0){\includegraphics[width=9cm]{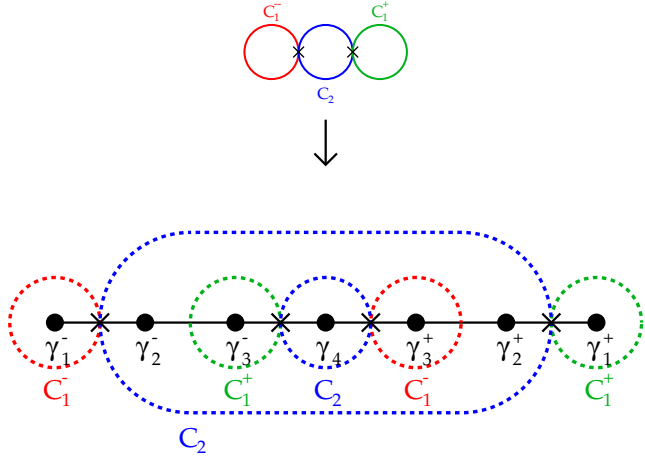}}}
\end{picture}
\end{center}
\caption[x]{\footnotesize Embedding of $A_3 \rightarrow A_7$ at an
  ordinary double point singularity, giving a two-index symmetric
  representation as well as antisymmetric representation
  ( ${\tiny\yng(2)} + {\tiny\yng(1,1)}$ ).}
\label{f:a3-a7}
\end{figure}

The explicit singularity resolution computed in Section
\ref{sec:ordinary-double-appendix} is depicted graphically in
Figure~\ref{f:a3-a7}, for a particular choice of relative orientation
of the $A_3$ curves in the two branes.
For this choice of orientation, the representation given is the
symmetric + antisymmetric representation.  This can be seen by
computing the Dynkin weight of the curve $\gamma_1^-+ \gamma_2^-+
\gamma_3^-+ \gamma_4+\gamma_3^+$.  The only nonzero inner product of this curve
with $C_1^\pm, C_2$ is
\begin{equation}
(\gamma_1^-+ \gamma_2^-+
\gamma_3^-+ \gamma_4+\gamma_3^+) \cdot C_1^-= -2 \,.
\label{eq:}
\end{equation}
The resulting Dynkin weight $[-2, 0, 0]$ occurs in the (conjugate of
the) symmetric representation, and not in the adjoint, so this
embedding corresponds to the matter representation
( ${\tiny\yng(2)} + {\tiny\yng(1,1)}$ ).

One simple class of global models that contain ordinary double point
singularities  is the set of models where an $A_N$
singularity is wrapped on a divisor class $\sigma$ that has a
self-intersection, but which can be continuously deformed into a
smooth divisor class without changing the gauge group of the theory.
In this case, the self-intersection is expected to generically be of
the type that gives an adjoint representation, since there is no
reason to expect the type of matter to change discontinuously as the
divisor becomes singular.  We give an example of such a configuration
in Section \ref{sec:6D}.  In some cases, however, a more complicated
global Weierstrass model can give self-intersections that produce
symmetric and antisymmetric matter fields.  The presentation of an
explicit example of such a configuration is left for the future work.

\subsection{Group theory of novel matter representations}

The range of possible codimension two singularities in F-theory is
very large, and provides an inviting territory for exploration.  One
guide in exploring this space is the set of matter representations
that may be expected to arise from F-theory singularity structures
based on analysis of low-energy theories.
As we discuss in more detail in Section \ref{sec:6D}, a systematic
analysis of $SU(N)$ matter representations in 6D supergravity theories
without tensor multiplets in \cite{0} identified a number of
representations that may arise in F-theory constructions.  In this
section we discuss the group theory aspect of how two of these
representations may arise.  Identification of local and global models
for singularity structures realizing these matter representations is
left for the future.

The two representations we focus on here are the 4-index antisymmetric
($\Lambda^4$) representation of $SU(8)$ with Young diagram
 ${\tiny\yng(1,1,1,1)}$  and the ``box'' representation of $SU(4)$ with
Young diagram  ${\tiny\yng(2,2)}$ 

\subsubsection{4-index antisymmetric representation of $SU(8)$}

To realize a representation $R$ of a group $G$  through the Katz--Vafa
analysis, $G$ must embed into a group $G'$ of one rank higher,
and the representation $R$ must appear in the decomposition of the
adjoint of $G'$ under $G\subset G'$.  At first appearance,
this seems difficult for the $\Lambda^4$ representation of $SU(8)$.
There is a natural embedding of $A_7$ into $E_8$ 
associated with the
obvious embedding of Dynkin diagrams 
\begin{center}
\begin{picture}(200,50)(- 100,- 35)
\put(0,0){\makebox(0,0){$\longrightarrow$}}
\put(-90,-5){\line( 1, 0){60}}
\put(30,-5){\line( 1, 0){60}}
\multiput(- 90,-5)(10,0){7}{\circle*{4}}
\multiput(30,-5)(10,0){7}{\circle*{4}}
\put(50,5){\circle{4}}
\multiput(50,-5)(0,4){3}{\line(0,1){2}}
\put(-60,-20){\makebox(0,0){$A_7$}}
\put(60,-20){\makebox(0,0){$E_8$}}
\end{picture}
\end{center}
Under this embedding, the adjoint of $E_8$ decomposes as
\cite{Slansky}
\begin{equation}
{\bf 248} ({\rm Adj}) \rightarrow {\bf 63} ({\rm Adj})
+{\bf 1} 
+  {\bf 28} \left( \;{\tiny\yng(1,1)}\; \right)
+  {\bf \bar{28}}
+ \left[ {\bf 8} \left( \;{\tiny\yng(1)}\; \right)
+  {\bf \bar{8}}
+ {\bf 56} \left( \;{\tiny\yng(1,1,1)}\;\right) + {\bf \bar{56}}  \right]\,.
\label{eq:78-1}
\end{equation}
The appearance of the $\Lambda^3$ representation, corresponding to
Dynkin indices $[0, 0, 1, 0, 0, 0, 0]$ is clear from the geometry
associated with the Dynkin diagram embedding depicted above.  The
$\P^1$ associated with the extra (empty) circle in the $E_8$ Dynkin
diagram has inner product 1 with the $\P^1$ associated with the third
root in the $A_7$ diagram, giving the Dynkin weight $[0, 0, 1, 0, 0,
  0, 0]$.

In addition to the above embedding, however, there is a second, inequivalent
embedding of $A_7 \subset E_8$ \cite{Dynkin,Oguiso-Shioda}.  
This alternate embedding can be understood through a
sequence of maximal subgroup embeddings $A_7 \subset E_7 \subset E_8$.
The form of the embedding $A_7 \subset E_7$ can be understood through
extended Dynkin diagrams.  In general \cite{BDS}, a maximal subgroup $H \subset
G$ of simple Lie algebras is associated with an embedding of the
Dynkin diagram of $H$ into the {\it extended} Dynkin diagram of
$G$.  The embedding of $A_7$ into the
extended Dynkin diagram of $E_7$ is depicted as (denoting the extra
node extending the $E_7$ with an ``x'')
\begin{center}
\begin{picture}(200,50)(- 100,- 35)
\put(0,0){\makebox(0,0){$\longrightarrow$}}
\put(-90,-5){\line( 1, 0){60}}
\put(30,-5){\line( 1, 0){60}}
\multiput(- 90,-5)(10,0){7}{\circle*{4}}
\multiput(30,-5)(10,0){7}{\circle*{4}}
\put(60,5){\circle{4}}
\multiput(60,-5)(0,4){3}{\line(0,1){2}}
\put(-60,-20){\makebox(0,0){$A_7$}}
\put(60,-20){\makebox(0,0){$\hat{E}_7$}}
\put(30,-12){\makebox(0,0){x}}
\end{picture}
\end{center}
The diagram suggests that the decomposition of the $E_7$ adjoint will
include a state in an $A_7$ representation with a Dynkin weight of
$[0, 0, 0, 1, 0, 0, 0]$, which is the desired highest weight of the
$\Lambda^4$ representation.  Indeed, under this embedding
of $A_7 \rightarrow E_7$ the adjoint decomposes as
\begin{equation}
{\bf 133} ({\rm Adj}) \rightarrow
{\bf 63} ({\rm Adj}) + {\bf 70}\left( \;{\tiny\yng(1,1,1,1)}\; \right) \,.
\end{equation}
Using this embedding to further
embed $A_7 \subset E_7 \subset E_8$ gives the decomposition of the
adjoint of $E_8$
\begin{equation}
{\bf 248} ({\rm Adj}) \rightarrow {\bf 63} ({\rm Adj})
+{\bf 1} 
+  {\bf 28} \left( \;{\tiny\yng(1,1)}\; \right)
+  {\bf \bar{28}}
+ \left[{\bf 1} + {\bf \bar{1}}  +{\bf  28}
+  {\bf \bar{28}}
+ {\bf 70} \left( \;{\tiny\yng(1,1,1,1)}\; \right)  \right]\,.
\label{eq:78-2}
\end{equation}
So under this embedding, the adjoint of $E_8$ decomposes in a way that
gives the $\Lambda^4$ representation of $A_7$.
Note that the representation content in brackets in \eq{eq:78-1},
\eq{eq:78-2}, giving the difference in content between the two
decompositions, is given in the two cases by
\begin{equation}
\Lambda^1 + \Lambda^3 + \Lambda^5 + \Lambda^7 \;\; \; {\rm vs.} \; \; \;
\Lambda^0 + \Lambda^2 + \Lambda^4 + \Lambda^6 + \Lambda^8  \,.
\end{equation}
This is precisely the difference in representation content between the
two spinor representations of $SO(16)$ when decomposed under $SU(8)$.

As we discuss further in Section \ref{sec:6D}, we anticipate that a
further analysis of global 6D models with $SU(8)$ gauge group will
provide Weierstrass forms that locally contain singularities
giving rise to matter in the $\Lambda^4$ representation of $SU(8)$.
The group theory structure just described is one natural way in which
this may occur.

\subsubsection{Box representation of $SU(4)$}
\label{sec:box}

Now let us consider the ``box'' representation of $SU(4)$.  In terms
of Dynkin indices, this representation is
\begin{equation}
{\tiny\yng(2,2)}({\bf 20'})  \;\leftrightarrow \;
[0,  2, 0] \,.
\label{eq:box}
\end{equation}
This representation does not appear in the decomposition of the
adjoint of any rank one gauge group enhancement.  Since the genus
\eq{eq:genus} of the representation is nonzero, we expect this
representation to arise from a Weierstrass singularity where the curve
on the base supporting the singularity locus is itself singular.  As
in the ordinary double point giving matter in the adjoint and
symmetric representations of $SU(N)$ through the embedding $A_{N
  -1}\rightarrow A_{N -1} \times A_{N -1} \rightarrow A_{2 N -1}$, we
look for a similar multiple embedding that may give rise to the
representation \eq{eq:box}.  As in the previous example,
such an embedding can be realized through
an embedding of $A_3 \times A_3$ into the extended Dynkin
diagram for $D_6$
\begin{center}
\begin{picture}(200,50)(- 100,- 35)
\put(0,0){\makebox(0,0){$\longrightarrow$}}
\put(-60,-5){\line( 1, 0){20}}
\put(40,-5){\line( 1, 0){40}}
\multiput(- 60,-5)(10,0){3}{\circle*{4}}
\multiput(40,-5)(10,0){2}{\circle*{4}}
\multiput(70,-5)(10,0){2}{\circle*{4}}
\put(50,5){\circle*{4}}
\put(70,5){\circle*{4}}
\put(60,-5){\circle{4}}
\put(50,-5){\line(0,1){10}}
\put(70,-5){\line(0,1){10}}
\put(-50,-20){\makebox(0,0){$A_3$}}
\put(60,-20){\makebox(0,0){$\hat{D}_6$}}
\put(40,-12){\makebox(0,0){x}}
\end{picture}
\end{center}
Under this embedding of $A_3 \rightarrow A_3 \times A_3 \rightarrow
D_6$ the adjoint of $D_6$ decomposes as
\begin{equation}
{\bf  66}  = 3 \times {\bf 15} + {\bf 1} + {\bf 20'} \,.
\end{equation}
The $D_6$ group can be further embedded in $D_7$ or $E_7$ giving a
rank one enhancement.  In either case, the box representation appears
in the decomposition of the adjoint.  As discussed further in Section
\ref{sec:6D}, we expect that a further analysis of 6D Weierstrass
models for $A_3$ on a singular curve of arithmetic genus 3 on $\P^2$
will give a global model with a local singularity type giving matter
in the box representation of $SU(4)$; the group theory mechanism just
described provides one natural way in which this may occur.

\section{Systematic analysis of Weierstrass models}
\label{sec:global}

We now perform a systematic analysis of Weierstrass models for $SU(N)$
gauge groups on a general F-theory base.  Thus we are looking for an
$A_{N -1}$ ($I_N$) Kodaira type singularity on a codimension  one
space described by a divisor
$\{\sigma=0\}$.  We assume in the analysis
that $\{\sigma=0\}$ is nonsingular, so that any ring of local functions $R_\sigma$ on
a sufficiently small open subset of
$\{\sigma=0\}$ is a unique factorization domain (UFD).  We comment on
extensions of this analysis to singular divisors $\{\sigma=0\}$ at various points in
the discussion.

The idea of this analysis is to use the Kodaira conditions on the form
of the singularity to determine the form of the coefficients $f, g$ in
the Weierstrass form for a fairly general class of models.  A related
analysis is carried out in \cite{Morrison-sn} using
the Tate form for
various gauge groups\footnote{The results of \cite{Morrison-sn} are 
complementary to the ones derived here, and include the case of 
$SU(N)$ for large $N$.}.  Here we
primarily use Weierstrass form since the counting of degrees of
freedom is clearest in this language.  The goal of the analysis here
is to follow various branches of the conditions on the discriminant
realized by a type $I_N$ singularity to identify models with matter
content associated with various local singularity types such as those
identified
in the previous section.

We begin with the Weierstrass form
\begin{equation}
y^2 = x^3 + fx + g \,.
\label{eq:Weierstrass-global}
\end{equation}
Here $f \in -4K$, $g \in -6K$ where $K$ is the canonical class on the
base $B$.
We expand
\begin{equation}
f=  \sum_{i}f_i \sigma^i \,,\qquad 
g=  \sum_i g_i \sigma^i \,.
\label{eq:fg-expansion}
\end{equation}
where as above, $\{\sigma=0\}$ is the codimension one locus on the base $B$
carrying the $A_{N -1}$ singularity.  For this general analysis we
leave the dimension of the base and degree of $\sigma$ unfixed.  In the
following sections we specialize to the cases where the base is a
complex surface (6D space-time theories) or complex 3-fold (4D
space-time theories).  In most situations we can consider $f_i, g_i$ as
polynomials in local coordinates $s, t$ (or $s, t, u$ for 4D theories)
on the base, with degrees that will depend on the particular
situation.  If we are working with an elliptically-fibered Calabi-Yau
$d$-fold over the base ${\P}^{d-1}$, and the degree of $\sigma$
is $b$ then the degrees of $f_i, g_i$ are
\begin{equation}
[f_i] = 4d-bi, \;\;\;\;\;[g_i] = 6d-bi \,.
\end{equation}
For example, for 6D theories with no tensor multiplets (the case
studied from the low-energy point of view in \cite{0}), the dimension
is $d = 3$, and $B =\P^2$, so for an $SU(N)$ group associated with a
singularity on a divisor class of degree $b = 1$, $f_0$ is a
polynomial in $s$ of degree 12, $f_1$ has degree 11, etc.
Note that since $f, g$ are really sections of line bundles, they  can
generally only be treated as functions locally.

The discriminant describing the total singularity locus is
\begin{equation}
\Delta = 4 f^3 + 27 g^2\,.
\label{eq:discriminant-2}
\end{equation}
We can expand the discriminant in powers of $\sigma$,
\begin{equation}
\Delta = \sum_{i}\Delta_i \sigma^i \,.
\end{equation}
For an $I_N$ singularity type we must have $\Delta_i = 0$ for $i < N$.
For each power of $\sigma$, the condition that $\Delta_i$
vanish imposes various algebraic conditions on the coefficients $f_i,
g_i$.  These conditions can be derived by a straightforward algebraic
analysis (some of which also appears in \cite{Morrison-sn}).  

For local functions $\Phi$ and $\Psi$ defined on an open set of the base
$B$, we use the notation
\begin{equation}\label{eq:notation}
\Phi \sim \Psi
\end{equation}
to indicate that $\Phi$ and $\Psi$ have identical restrictions
to $\{\sigma=0\}$, i.e., $\Phi|_{\{\sigma=0\}}=\Psi|_{\{\sigma=0\}}$.
Equivalently, $\Phi$ and $\Psi$ differ by a multiple of $\sigma$,
i.e., $\Phi = \Psi + {\cal O}(\sigma)$.

We proceed by systematically imposing the condition that the
discriminant (\ref{eq:discriminant-2}) vanish at each order in a
fashion compatible with an $A_{N -1}$ singularity on $\{\sigma=0\}$.
\vspace*{0.05in}

\noindent $\Delta_0 = 0$:

The leading term in $\Delta$ is
\begin{equation}
\Delta_0 = 4 f_0^3 + 27 g_0^2 \,.
\end{equation}
For this to vanish in a fashion compatible with an $A_{N -1}$
singularity, we must be able to locally express $f_0, g_0$ in terms of some
$\phiorig$ by
\begin{eqnarray}
f_0 & \sim &  -\frac1{48} \phiorig^2 \label{eq:dorig}\\
g_0 & \sim &  \frac1{864} \phiorig^3 \nonumber 
\end{eqnarray}
Moreover, when $N\ge3$, $\phiorig$ has a square root (locally), and
we can rewrite this condition as
\begin{eqnarray}
f_0 & \sim &  -\frac1{48} \phit^4 \label{eq:d0}\\
g_0 & \sim &  \frac1{864} \phit^6 \nonumber 
\end{eqnarray}
The condition that $f_0|_{\{\sigma=0\}} = x^2$ for some $x \in R_\sigma$ follows from the condition
that the ring of local functions on sufficiently small open subsets of
the variety defined by $\{\sigma=0\}$ is a
unique factorization domain (so each factor of $g_0|_{\{\sigma=0\}}$ must appear an
even number of times in $(f_0|_{\{\sigma=0\}})^3$); the local function $x$
on the divisor $\{\sigma=0\}$ can then be ``lifted'' to a function $X$
on an open subset of $B$ such that $X|_{\{\sigma=0\}}=x$.
  Note that the existence of $X$
is definitely only a local property in general: \cite{Morrison-sn} has
an explicit example that shows that it may not be possible to find $X$ (or
its square root when that is appropriate) globally.
The  condition that $X$ is itself a
square modulo $\sigma$ follows from  the ``split'' form of the singularity in the Tate algorithm \cite{Tate,Bershadsky-all} for
determining the Kodaira singularity type from the Weierstrass 
form\footnote{When $N=2$, there is no split form and no monodromy,
and we cannot conclude that $X$ is a square modulo $\sigma$.}.
This condition can be seen explicitly in the $A_3$ and $A_5$ examples
described in Sections \ref{sec:a3}
and \ref{sec:a5}.  In those
cases, $f_0$ is proportional to $s^4$, modulo $\sigma$.  If $s^4$ in these situations
were replaced with $s^2$, the exceptional curve in the first chart
would be defined by $y^2 = sx^2$, and would not factorize into
$C_1^\pm$, so that the resulting gauge group would be the symplectic
group $Sp(N)$ instead of $SU(N)$.
The numerical coefficients in (\ref{eq:dorig}) and
(\ref{eq:d0}) are chosen to simplify
parts of the algebra in other places and to match with other papers
including \cite{Morrison-sn}
\footnote{For reference, we give a dictionary relating the variables
  used here to analogous variables used in \cite{Morrison-sn}.  The variables
$(\phit,\phif,\phisa,\dots,\phie,\phis,\dots)$
in this paper correspond to the variables
$(s_0,u_1,u_2,\dots,t_1,t_2,\dots)$
in
\cite{Morrison-sn}.  Note that $\mu$ from \cite{Morrison-sn} must be
set equal to $1$ to match this paper.}.
\vspace*{0.05in}

We phrase the arguments of this section in terms of quantities such
as $\phi_0$ that are in general only locally defined
functions.  However, in some key examples (such as the ones at the
beginning of Section~\ref{sec:6D}) it is known that these quantities
are actually globally defined on $B$.  In those cases, we are easily
able to count parameters in the construction by considering the degrees
of these globally defined objects.

\noindent $\Delta_1 = 0$:

In light of (\ref{eq:dorig}), we now replace $f_0$ and $g_0$ by
$-\phiorig^2/48$ and $\phiorig^3/864$, respectively.  This may produce
additional contributions to $f, g$ at higher order in $\sigma$, since
for example the original $f_0$ was only equal to $-\phiorig^2/48$ up to
terms of order $\sigma$.  Such additional contributions can be
absorbed by redefining the coefficients $f_i$ and $g_i$ from
(\ref{eq:fg-expansion}) accordingly.  The coefficient of the leading
term in the discriminant then becomes
\begin{equation}
\Delta_1 =\frac{1}{192}  \left(12 \phiorig^3 g_1 + \phiorig^4 f_1 \right)
 \,.
\end{equation}
This vanishes exactly when
\begin{eqnarray}
g_1 & = &  -\frac1{12}\phiorig f_1 \,. \label{eq:d1}
\end{eqnarray}
A similar term must be removed from $g_i$ at each order (this can be
seen just from the terms $g_0g_i, f_0^2 f_i$ in the discriminant; a
more general explanation for this structure is described at the end of
this section), so we generally
define
\begin{equation}
\tilde{g_i} = g_i +\frac1{12} \phiorig f_i 
\label{eq:gt}
\end{equation}

\noindent $\Delta_2 = 0$:

After imposing (\ref{eq:d0}) (as a substitution)
and (\ref{eq:d1}), the coefficient of the
next term in the
discriminant is
\begin{equation}
\Delta_2 =\frac{1}{16}  \left(\phiorig^3 \tilde{g}_2  - \phiorig^2 f_1^2 \right)
 \,.
 \label{eq:delta2}
\end{equation}
At this stage, we also impose the condition $\phiorig=\phit^2$ to guarantee
$SU(N)$ gauge symmetry, so that the next term in the discriminant becomes
\begin{equation}
\Delta_2 =\frac{1}{16}  \left(\phit^6 \tilde{g}_2  - \phit^4 f_1^2 \right)
 \,.
 \label{eq:delta2bis}
\end{equation}

For (\ref{eq:delta2bis}) to vanish in our UFD, $f_1|_{\{\sigma=0\}}$ 
must be divisible by $\phit|_{\{\sigma=0\}}$, so there is a locally
defined function $\psi_1$ such that
\begin{equation}
f_1  \sim    \frac12 \phit \phie .\label{eq:d2}
\end{equation}
We replace $f_1$ by $\frac12\phit\phie$ and adjust coefficients
accordingly; we can then solve $\Delta_2=0$ for $\tilde{g}_2$, obtaining:
\begin{equation}
\tilde{g}_2 =  \frac14 \phie^2. \label{eq:d2bis}
\end{equation}
(Note from (\ref{eq:gt}) that
this last equation is equivalent to 
$g_2=\frac14\phie^2-\frac1{12}\phit^2f_2$.)

\vspace*{0.1in}

\noindent ${\bf SU(4)}$  ($\Delta_3 = 0$):

At the next order in $\sigma$
the coefficient in the discriminant is
\begin{equation}
\Delta_3 =\frac{1}{16}  \left(\phit^6 \tilde{g}_3-\phit^3 \phie^3  -
\phit^5 \phie f_2 \right)
 \,.
\label{eq:d3}
\end{equation}
We see that in order for $\Delta_3$ to vanish along $\{\sigma=0\}$,
 $\phie|_{\{\sigma=0\}}$ must be divisible by $\phit|_{\{\sigma= 0\}}$.
Thus, there must exist a locally defined function $\phi_1$ such that
\begin{equation}
 \phie  \sim    -\frac13 \phit \phif .
\end{equation}
We replace $\phie$ by $-\frac13\phit\phif$ and adjust coefficients
accordingly; we can then solve $\Delta_3=0$ for $\tilde{g}_3$,
obtaining:
\begin{equation}
\tilde{g}_3 = -\frac13\phif f_2 -\frac1{27}\phif^3 \label{eq:g2}
\end{equation}
(This last equation is equivalent to
$g_3 = -\frac1{12}\phit^2f_3-\frac13\phif f_2 -\frac1{27}\phif^3$.)
Again,  a term such as the first term on the RHS of (\ref{eq:g2})
will arise for each $\tilde{g}_i$, so we
define
\begin{equation}
\hat{g_i} = \tilde{g}_i + \frac13\phif f_{i-1}
\end{equation}
and the latter condition (\ref{eq:g2}) is just $\hat{g}_3 = -\phif^3/27$.
It is also convenient to define $\hat{f}_2 = f_2+\frac13\phif^2$.

We have now arranged a theory with an $SU(4)$ local factor in the gauge group.
  The construction is completely general, given our assumption
about $\{\sigma=0\}$ being nonsingular.
Making the substitutions above, adjusting coefficients,
and expanding $f, g,$ and $\Delta$ 
we have
\begin{align}
f & =  -\frac1{48} \phit^4 -\frac16 \phit^2 \phif \sigma + f_2 \sigma^2 + 
f_3\sigma^3+f_4\sigma^4+{\cal O}(\sigma^5) \label{eq:4-f}\\
g & =   \frac1{864} \phit^6+\frac1{72} \phit^4 \phif \sigma
+ (\frac1{36} \phit^2 \phif^2 -\frac1{12}\phit^2 f_2) \sigma^2 +
( -\frac1{12}\phit^2 f_3 -\frac13  \phif f_2 -\frac1{27} \phif^3)
\sigma^3 
\label{eq:4-g}\\
&\quad + {g}_4 \sigma^4 + {\cal O} (\sigma^5)
 \notag \\
\Delta & = 
\frac{1}{16}\phit^4(- \hat{f}_2^2 + 
\phit^2 \hat{g}_4)  \sigma^4 +{\cal O} (\sigma^5)  \label{eq:d4}
\end{align}

We see that at a generic point on the curve $\{\sigma=0\}$ the
singularity type is $I_4$, with vanishing degrees of $f, g, \Delta$ of
0, 0, $4$, corresponding to an $A_3$ singularity giving a $SU(4)$
gauge group.  At the roots of $\phit$, the vanishing degrees become 2,
3, 6, corresponding to a $D_4$ singularity, giving a two-index
antisymmetric ($\Lambda^2$) matter representation.  The remaining part
of the leading component of the discriminant, $\tilde{\Delta}_4 =
\Delta_4/\phit^4 = (- \hat{f}_2^2 + \phit^2 \hat{g}_4)/16$, is of
degree $8d -4b$.  For generic choices of the coefficients of the other
functions $\phif, f_2, \ldots$, the roots of $\tilde{\Delta}_4$ will
correspond to an enhancement to $A_4$, giving matter in the
fundamental representation of $SU(4)$.  For non-generic choices of the
functions $f_2, \phif$, there can be enhanced singularities.  In
particular if $f_2$ and $\phit$ share a root the degree of vanishing
of $f$ is enhanced to 3.  The following table shows the possibilities
for enhanced singularities
\begin{center}
\begin{tabular}{| c | c | c | c | c | c | c|}
\hline
Label & Root & ${f}$ & ${g}$ & ${\Delta}$ & Singularity & G/Rep.\\
\hline
$4_0$ & generic & 0 & 0 & 4 & $A_3$ & $SU(4)$\\
\hline
&&&&&&\\[-6pt]
$4_a$ & $ \tilde{\Delta}_4 = 0$ &   0 &  0 &  5 & $A_4$ & {\tiny\yng(1)}\\
$4_b$ & $ \phit = 0$ &  2 & 3 & 6 & $D_4$ & ${\tiny\yng(1,1)}$
($\Lambda^2$)\\[4pt]
\hline
&&&&&&\\[-5pt]
$4_c$ & $ \phit = 0, f_2 = 0$ &  3 & 3 & 6 & $D_4$ & ${\tiny\yng(1,1)}$\\[8pt]
$4_d$ & $ \phit = f_2 = \phif = 0$  & 3 & 4 & 8 & $E_6$ & 
$\left[ \;{\tiny\yng(1,1)} + 2 \; {\tiny\yng(1)}\;\right]$\\[8pt]
\hline
\end{tabular}
\end{center}
The explicit local resolution of singularity type $4_b$ with
$A_3$ enhancement to $D_4$ is that described in Section \ref{sec:a3},
with details in Section \ref{sec:a3-appendix} of the Appendix.
Replacing $\phit \rightarrow 2s, f_2 \rightarrow -1,$ and for
simplicity $\phif \rightarrow 0$ (which does not affect the
singularity), $f, g$ from \eq{eq:4-f} and \eq{eq:4-g} take precisely
the forms \eq{eq:a3-fg} used in that analysis.
In the last case ($4_d$)
a more exotic singularity appears but
no new matter representations arise.  The brackets in
the table indicate that we have not explicitly resolved the
singularity, but the matter content is uniquely determined by the 6D
anomaly cancellation conditions, as we discuss in the following
section.  
\vspace*{0.1in}

\noindent ${\bf SU(5)}$  ($\Delta_4 = 0$):

The vanishing
of the leading term in
(\ref{eq:d4}) requires that $\hat{f}_2|_{\{\sigma=0\}}$ be
divisible by $\phit|_{\{\sigma=0\}}$. Thus, in this case
there exists a locally defined function $\phis$ such that
\begin{equation}
\hat{f}_2 \sim \frac12 \phit \phis .
\end{equation}
We replace $\hat{f}_2$ by $\frac12\phit\phis$ and adjust  coefficients
accordingly; we can then solve $\Delta_4=0$ for $\hat{g}_4$,
obtaining:
\begin{equation}
\hat{g}_4  =   \frac{1}{4} \phis^2 \,.
\end{equation}
(In other words, $f_2$ has been replaced by
$\frac12 \phit \phis-\frac13 \phif^2$
and $g_4= \frac{1}{4} \phis^2 - \frac1{12} \phit^2f_4 - \frac13 \phif f_3$.)
We have now arranged a theory with an $SU(5)$ local factor in the gauge group
 (again completely general, assuming $\{\sigma=0\}$ is 
nonsingular).  Expanding $f, g,$ and $\Delta$ 
we have
\begin{align}
f & =  -\frac1{48} \phit^4 -\frac16 \phit^2 \phif \sigma + (\frac12 \phit \phis-\frac13 \phif^2) \sigma^2 + 
f_3\sigma^3+f_4\sigma^4+f_5\sigma^5+{\cal O}(\sigma^6) \label{eq:5-f}\\
g & =   \frac1{864} \phit^6+\frac1{72} \phit^4 \phif \sigma
+ (\frac1{18} \phit^2 \phif^2 -\frac1{24}\phit^3 \phis)
\sigma^2  \label{eq:5-g}\\
& \quad+
( -\frac1{12}\phit^2 f_3 -\frac16 \phit \phif \phis + \frac{2}{27} \phif^3  ) \sigma^3 \nonumber\\
&\quad 
+ (\frac{1}{4} \phis^2 - \frac1{12} \phit^2f_4 - \frac13 \phif f_3)
 \sigma^4 + g_5\sigma^5+{\cal O} (\sigma^6)
 \notag \\
\Delta & = 
\frac1{16}\phit^4(
\phit^2 \hat{g}_5-\phit\phis f_3+\phif\phis^2)  \sigma^5 +{\cal O} (\sigma^6)  \label{eq:d5}
\end{align}

The range of possible singularities is similar to that encountered in
the $SU(4)$ case above.  At the roots of $\phit$
the singularity type is enhanced to $D_5$, and the roots of
the remaining $\tilde{\Delta}_4 = \Delta_4/\phit^4$ give $A_5$
singularities.  There are also various enhanced singularities for
non-generic configurations, but no new matter representations are
possible.  We again summarize the possible singularity types in the
following table
\begin{center}
\begin{tabular}{| c | c | c | c | c | c | c|}
\hline
Label & Root & ${f}$ & ${g}$ & ${\Delta}$ & Singularity & G/Rep.\\
\hline
$5_0$ & generic & 0 & 0 & 5 & $A_4$ & $SU(5)$\\
\hline
&&&&&&\\[-8pt]
$5_a$ & $ \tilde{\Delta}_4 = 0$ &   0 &  0 &  5 & $A_5$ & {\tiny\yng(1)}\\
$5_b$ & $ \phit = 0$ &  2 & 3 & 7 & $D_5$ & ${\tiny\yng(1,1)}$ ($\Lambda^2$)\\[4pt]
\hline
&&&&&&\\[-6pt]
$5_c$ & $ \phit = \phif = 0$ &  3 & 4 & 8 & $ E_6$ & 
$\left[ \;{\tiny\yng(1,1)} + {\tiny\yng(1)} \;\right]$\\[6pt]
$5_d$ & $ \phit = \phis = 0$  &  2 & 3 & 8 & $D_6$ &
$\left[ \;{\tiny\yng(1,1)} + {\tiny\yng(1)}\;\right]$\\[6pt]
$5_e$ & $ \phit =  \phif = \phis = 0$  & 3 & 5 & 9 & $E_7$ & 
$\left[ \;{\tiny\yng(1,1)} + 2\;{\tiny\yng(1)}\;\right] $\\[6pt]
\hline
\end{tabular}
\end{center}
\vspace*{0.1in}

\noindent ${\bf SU(6)}$  ($\Delta_5 = 0$):

The analysis becomes more interesting at the next order.  Using the
above conditions the leading order term in the discriminant is
\begin{equation}
\Delta_5 = \frac1{16}\phit^4 (\phif \phis^2 - \phit \phis f_3 +
\phit^2 \hat{g}_5) \,.
\label{eq:de5}
\end{equation}
From this it follows that each root of $\phit|_{\{\sigma=0\}}$ must either divide
$\phif|_{\{\sigma=0\}}$ or $\phis|_{\{\sigma=0\}}$.    We can
find locally defined functions $\alpha$ and $\beta$ such that
\begin{equation}
\phit \sim \alpha \beta \,,
\label{eq:pt-ab}
\end{equation}
where $\alpha|_{\{\sigma=0\}}$ is the greatest common divisor of
$\phit|_{\{\sigma=0\}}$ and $ \phis|_{\{\sigma=0\}}$.  There must then also be locally
defined functions $\phisa$ and $\phifb$ such that
\begin{eqnarray}
\phis & \sim &  -\frac13\alpha \phisa \label{eq:psf-ab}\\
\phif & \sim &  \beta \phifb \,. \label{eq:pf}
\end{eqnarray}
Note that by construction, $\beta|_{\{\sigma=0\}}$ and $\phisa|_{\{\sigma=0\}}$
are relatively prime.

We make all of the corresponding substitutions and adjust coefficients;
then  (\ref{eq:de5}) becomes:
\begin{equation} \label{eq:de5bis}
\Delta_5 =
\frac1{48}\alpha^6\beta^5\left(\phisa (f_3+\frac13\phifb\phisa)
+ 3\beta \hat{g}_5 \right) \,.
\end{equation}
In order for this to vanish we then must have
$(f_3+\frac13\phifb\phisa)|_{\{\sigma=0\}}$ divisible by 
$-3\beta|_{\{\sigma=0\}}$
and $\hat{g}_5|_{\{\sigma=0\}}$ divisible by $\phisa|_{\{\sigma=0\}}$,
with identical quotients.  That is, there must exist a
locally defined function $\lambda$ such that
\begin{eqnarray}
f_3 & \sim &  -\frac13\phifb \phisa -3 \beta\lambda \label{eq:f3e}\\
\hat{g}_5 & \sim &     \phisa \lambda \,.
\end{eqnarray}
The second relation can also be written as
\begin{equation}
\begin{split}
g_5 &\sim -\frac1{12} \phit^2f_5 -\frac13\phif f_4 + \phisa \lambda\\
&\sim -\frac1{12} \alpha^2\beta^2f_5 -\frac13\beta\phifb f_4 + \phisa \lambda
\end{split}
\end{equation}

The possible singularities are now
\begin{center}
\begin{tabular}{| c | c | c | c | c | c | c|}
\hline
Label & Root & ${f}$ & ${g}$ & ${\Delta}$ & Singularity & G/Rep.\\
\hline
$6_0$ & generic & 0 & 0 & 6 & $A_5$ & $SU(6)$\\
\hline
&&&&&&\\[-8pt]
$6_a$ & $ \tilde{\Delta}_6 = 0$ &   0 &  0 &  7 & $A_6$ & {\tiny\yng(1)}\\[6pt]
$6_b$ & $ \alpha = 0$ &  2 & 3 & 8 & $D_6$ & ${\tiny\yng(1,1)}$ ($\Lambda^2$)\\[4pt]
\hline 
&&&&&&\\[-6pt]
$6_c$ & $ \beta = 0$ & 3 & 4 & 8 & $E_6$ & $\frac{1}{2}  \;
{\tiny\yng(1,1,1)}\; (\Lambda^3$)\\[8pt]
$6_d$ & $ \alpha = \beta = 0$  &  3 & 5 & 9 & $E_7$ & 
$\left[\frac{1}{2}\;{\tiny\yng(1,1,1)} + {\tiny\yng(1,1)}\;\right]$\\[8pt]
$6_e$ & $ \beta =  \phifb = 0$  & 4 & 4 & 8 & $E_6$ & 
$\left[\frac{1}{2} \;{\tiny\yng(1,1,1)} \;\right]$\\[8pt]
$6_f$ & $  \alpha =  \phifb = 0$  & 3 & 5 & 9 & $E_7$ & 
$\left(\left[\frac{1}{2} \;{\tiny\yng(1,1,1)} + {\tiny\yng(1)} \;\right)
/ \; {\tiny\yng(1,1)}\;\right]$\\[8pt]
\hline
\end{tabular}
\end{center}
We see now the appearance of a 3-index antisymmetric matter field.
The singularity types $6_b$ and $6_c$  are precisely the enhancements
of $A_5$ to $D_6$ and $E_6$
analyzed locally in Section \ref{sec:a5}, with details in
Section \ref{sec:a5-appendix} of the Appendix.
To relate \eq{eq:5-f}, \eq{eq:5-g} to the local forms there we 
use
\eq{eq:pt-ab}, \eq{eq:psf-ab}
and
make
the replacements
\begin{equation}
(6_b): \;
\alpha \rightarrow s, \beta \rightarrow 2,
\phisa \rightarrow -6, \phifb \rightarrow 3/2,
\lambda \rightarrow 0, f_4
\rightarrow 0,
\end{equation}
\begin{equation}
(6_c): \;
\alpha \rightarrow 1, \beta \rightarrow  2 s,
\phisa \rightarrow -6, \phifb \rightarrow 3/2,
\lambda \rightarrow 0, f_4
\rightarrow 0 \,.
 \label{eq:6c-replacement}
\end{equation}
The replacement \eq{eq:6c-replacement} gives \eq{eq:phi-a5-e6} with
$\rho = s$.  Thus produces a half-hypermultiplet in the $\Lambda^3$
representation.  Two coincident roots of $\beta$ give $\rho = s^2$,
for a full hypermultiplet in the $\Lambda^3$ representation, as
discussed in Section \ref{sec:a5}.

It is interesting to note that the $6_b$ and $6_c$ singularities with
$D_6$ and $E_6$ enhancements are connected.  If we consider a $6_c$
branch with $\beta = 0$, we can continuously deform the coefficients of
the Weierstrass form so that the root $\beta$ coincides with a root of
$\phisa$.  At this point, the root
of $\phisa$ divides $\phit$, so in the decomposition \eq{eq:pt-ab},
\eq{eq:psf-ab} the simultaneous root of $\beta, \phisa$ becomes a root
of $\alpha, \phifb$,
giving a singularity of type $6_f$.  The root of $\alpha$ can then be deformed
independently of $\phifb$.  In six dimensions, this  deformation
transforms a combination of a half hypermultiplet in the $\Lambda^3$
representation and a hypermultiplet in the fundamental representation
into a single hypermultiplet in the $\Lambda^2$ representation.  This
novel phase transition is clear from the F-theory description but does
not have a simple description in the low-energy theory in terms of
Higgsing.  We describe an explicit example of a transition of this
kind in a specific 6D theory in the following section.
Note that the intermediate state in this transition associated with a
singularity of type $6_f$ involves a local enhancement $A_5 \subset
E_7$ with rank increase of more than  one.  This kind of transition
will be discussed further elsewhere.
\vspace*{0.1in}

\noindent ${\bf SU(7)}$  ($\Delta_6 = 0$):

At order 7, it becomes more difficult to identify the general
Weierstrass form.  Imposing the conditions above, the 6th order term
in the discriminant is
\begin{equation}
\Delta_6 =\frac{1}{16}\alpha^4 \beta^3 \left[
-\frac19 \beta \left( 
\phifb \phisa
- 9\beta\lambda
\right)^2
+\alpha^2 \left( \frac1{27} \phisa^3
+ \frac13\beta^2 \phisa f_4 +   \beta^3 \hat{g}_6\right) \right]
\label{eq:d6}
\end{equation}
We do not have a completely general form for the structure needed to make this
term vanish.  But there are two special cases in which we can carry
out the analysis
and guarantee the vanishing
of (\ref{eq:d6})

\vspace*{0.05in}

\noindent {\bf Case 7A}

\begin{eqnarray}
\beta & = & 1\\
\lambda & = &  \frac19\phifb \phisa -\frac16 \ola \alpha \label{eq:lambda}\\
\hat{g}_6 & = &  -\frac1{27}\phisa^3 +  \frac14\ola^2
  -\frac13 \phisa f_4 \,.
\end{eqnarray}
In this case the local singularities can appear as in the following
table
\begin{center}
\begin{tabular}{| c | c | c | c | c | c | c|}
\hline
Label & Root & ${f}$ & ${g}$ & ${\Delta}$ & Singularity & G/Rep.\\
\hline
$7_0$ & generic & 0 & 0 & 7 & $A_6$ & $SU(7)$\\
\hline
&&&&&&\\[-6pt]
$7_a$ & $ \tilde{\Delta}_7 = 0$ &   0 &  0 &  8 & $A_7$ & {\tiny\yng(1)}\\[4pt]
$7_b$ & $ \alpha = 0$ &  2 & 3 & 9 & $D_7$ & ${\tiny\yng(1,1)}$ ($\Lambda^2$)\\[4pt]
\hline
&&&&&&\\[-8pt]
$7_c$ & $ \alpha = \phifb = 0$ & 4 & 6 & 12 & $\star $ &  $\Delta T$\\[2pt]
\hline
\end{tabular}
\end{center}
In case $7_c$ the singularity of degrees 4, 6, 12 goes outside the
Kodaira list.  To resolve the singularity, the codimension two
singularity locus on the base must be blown up.  In six-dimensional
gravity theories this leads to the appearance of an additional tensor
multiplet.

\vspace*{0.05in}

\noindent {\bf Case 7B}

In general, for (\ref{eq:d6}) to vanish we must have 
$ (\alpha|_{\{\sigma=0\}})^2$
divisible by
$\beta|_{\{\sigma=0\}}$.
We can then write
\begin{equation}
\beta \sim  \gamma \delta^2
\label{eq:b-gd}
\end{equation}
for  appropriate locally defined functions $\gamma$ and $\delta$ such that
 $(\gamma \delta)|_{\{\sigma=0\}}$ is the GCD of $\alpha|_{\{\sigma=0\}}$
and $ \beta|_{\{\sigma=0\}}$.  We must then
have 
$ \alpha|_{\{\sigma=0\}}$
divisible by
$(\gamma|_{\{\sigma=0\}})^2 $
and furthermore we can decompose
\begin{eqnarray}
\alpha & \sim &  \gamma^2 \delta \alphabeta\\
\phifb & \sim &  \gamma \zeta \,.
\end{eqnarray}
for appropriate locally defined functions $\alphabeta$ and $\zeta$.
We can arrange for (\ref{eq:d6}) to vanish (case B)
if we make the assumption
that $\gamma = \alphabeta =1$, so that $\beta \sim \alpha^2$.
In this case the singularities that can arise are
\begin{center}
\begin{tabular}{| c | c | c | c | c | c | c|}
\hline
Label & Root & ${f}$ & ${g}$ & ${\Delta}$ & Singularity & G/Rep.\\
\hline
$7'_0$ & generic & 0 & 0 & 7 & $A_6$ & $SU(7)$\\
\hline
&&&&&&\\[-6pt]
$7'_a$ & $ \tilde{\Delta}_7 = 0$ &   0 &  0 &  8 & $A_7$ & {\tiny\yng(1)}\\[8pt]
$7'_b$ & $ \alpha = \beta = 0$ &  3 & 5 & 9 & $E_7$ & $\left[\;{\tiny\yng(1,1,1)} (\Lambda^3)\;\right]$\\[8pt]
\hline
&&&&&&\\[-8pt]
$7'_c$ & $ \alpha = \phifb = 0$ & 4 & 6 & 13 & $\star $ &  $\Delta T$\\[2pt]
\hline
\end{tabular}
\end{center}
The  singularities at $\alpha = \beta$ give rise to 3-index
antisymmetric matter representations of SU(7).

\vspace*{0.1in}

\noindent  ${\bf SU(8)}$ {\bf and beyond}

A complete treatment of all possible branches of the Weierstrass model
for $A_7$ and beyond would be very involved algebraically.  We do not
attempt a complete analysis but describe the generic structure of
Weierstrass models giving codimension one $A_{N -1}$ singularities for
$N \geq 8$.  To proceed further we need to get $\Delta_7$ to vanish.
This cannot be done in case B above since vanishing at order 8 given
the conditions imposed in that case would give a common root to
$\beta$ and $\phisa$, which is not possible since $\beta$ and $\phisa$
are relatively prime.  We can, however proceed to arbitrary order in
$N$ under the generic assumption that $\beta = 1$.  This corresponds
to case 7A above.  Note that all of the representations beyond the
fundamental and $\Lambda^2$ representations arose from situations
where $\beta \neq 1$.

First, we note that the condition $\beta = 1$ simplifies the algebra
at $SU(6)$ and beyond.  This condition sets $\alpha = \phit$ and
replaces \eq{eq:psf-ab} with
\begin{equation}
\phis \sim-\frac13\phit \phisa \,,
\end{equation}
and fixes $\phifb = \phif$.  Furthermore, \eq{eq:f3e} and
\eq{eq:lambda} become
\begin{equation}
f_3 \sim\frac12\phit\ola -\frac23\phif\phisa \,.
\end{equation}
We can proceed with the generic $A_{N -1}$ model by simply following
this pattern.  To get an $SU(8)$ model we substitute
\begin{equation}
\ola \sim-\frac13\pz  \phi_3 \,,
\end{equation}
and solve for $g_7$ \,.
To get an $SU(9)$ model we substitute
\begin{equation}
f_4 \sim \frac12\phit \psi_4-\frac23\phif \phi_3-\frac13 \phisa^2  \,,
\label{eq:f4}
\end{equation}
and solve for $g_8$ \,.
To get an $SU(10)$ model we substitute
\begin{equation}
\psi_4 \sim-\frac13\pz \phi_4 \,,
\label{eq:s4}
\end{equation}
and solve for $g_9$, etc.

A simple way of expressing the conditions being imposed is that the
leading terms in the expansions of $f, g$ can be written in the form
\begin{eqnarray}
f & = &  -\frac13 \Phi^2 + {\cal O}(\sigma^k)
\label{eq:expansion-f-even}\\
g + \frac13 \Phi f & = &  -\frac1{27} \Phi^3 + {\cal O}(\sigma^{2k})
\label{eq:expansion-g-even}\,
\end{eqnarray}
for $SU(2k)$, and
\begin{eqnarray}
f & = &  -\frac13 \Phi^2 + \frac12\sigma^k\phi_0\psi_k+ {\cal O}(\sigma^{k+1})
\label{eq:expansion-f-odd}\\
g + \frac13 \Phi f & = &  -\frac1{27} \Phi^3 + \frac14\sigma^{2k}\psi_k^2+ {\cal O}(\sigma^{2k+1})
\label{eq:expansion-g-odd}\,
\end{eqnarray}
for $SU(2k+1)$,
where
\begin{equation}
\Phi=\frac{1}{4} \phi_0^2 +\phi_1 \sigma +\phi_2\sigma^2+\phi_3\sigma^3 +
  \cdots + \phi_{k-1}\sigma^{k-1} \,.
\end{equation}
(This is the same form used in
 the inductive argument 
given in \cite{Morrison-sn}
for  $SU(N)$ with large $N$.)

In this way, we can find 
a systematic solution out to the point where there are no
more $g_i$'s for which to solve.  In the following section we describe
the details of how the analysis continues beyond this point for
a specific class of 6D models.

The numerical factors here, and the form of the equation, can be explained
by converting our Weierstrass equation (\ref{eq:Weierstrass-global})
to Tate form.  
Let $\Upsilon=\phi_1  +\phi_2\sigma+\phi_3\sigma^2 + \cdots
+ \phi_{k-1}\sigma^{k-2}$, so that $\Phi = \frac14\phi_0^2+\sigma\Upsilon$.
For $SU(2k)$, we convert to Tate form using the coordinate change 
\begin{eqnarray}
x & = & X+\frac13\Phi \\
y & = & Y + \frac12\phi_0X
\end{eqnarray}
giving an equation of the form
\begin{equation} \label{eq:Tate-even}
Y^2 + \phi_0 XY = X^3 + \sigma\Upsilon X^2 + \sigma^k FX + \sigma^{2k} G \,.
\end{equation}
Similarly, for $SU(2k+1)$, we convert to Tate form using the
coordinate change
\begin{eqnarray}
x &=& X + \frac13 \Phi \\
y &=& Y + \frac12\phi_0X + \frac12 \sigma^k\psi_k
\end{eqnarray}
giving an equation of the form
\begin{equation} \label{eq:Tate-odd}
Y^2 + \phi_0 XY + \sigma^k\psi_k Y = X^3 + \sigma\Upsilon X^2 + \sigma^{k+1}FX
+ \sigma^{2k+1}G\,.
\end{equation}

\section{6D supergravity without tensor fields}
\label{sec:6D}

We now use the general analysis of the previous section to describe a
particular class of 6D supergravity theories arising from F-theory.
We consider the class of 6D models with no tensor multiplets ($T = 0$)
and a gauge group having a nonabelian local factor $SU(N)$.  These theories
correspond to F-theory constructions on the base $\P^2$.

In \cite{0} theories of this kind were analyzed from the point of view
of the anomaly cancellation conditions in the low-energy theory.  A
complete list of all possible matter representations for each local gauge
group factor $SU(N)$ was constructed for theories with $T = 0$.  
From the point of view of the low-energy theory, each local $SU(N)$
factor is associated with an integer
$b \in\Z_+$ appearing in the anomaly polynomial and topological $BF^2$
couplings of the theory.  For theories with an F-theory realization,
$b$ is the degree of the divisor on $\P^2$ carrying the $SU(N)$ local factor.
For
small values of $b$,  anomaly analysis  of the 6D supergravity theories shows that
$N$ can range up to 24, and the
set of possible matter representations is strongly constrained.  For
larger values of $b$ the range of possible values of $N$ is more
restricted, but a wider range of possible matter representations is
compatible with the anomaly conditions.  We now recall from \cite{0}
the possible matter content for models with gauge group $SU(N)$ and
small values of $b$, and consider the explicit F-theory constructions
of such models.


\subsection{$SU(N)$ on curves of degree $b = 1$}

From anomaly cancellation alone, the complete set of possible matter
representations for an $SU(N)$ local factor with $b =1$ in a 6D ${\cal N} =
1$ supergravity theory is constrained to the following combinations of
matter fields (note that for $N = 3$ the antisymmetric $\Lambda^2$
representation is really the (conjugate of) the fundamental
representation while for $N = 2$ the fields denoted by this
representation are really uncharged.):

\begin{center}
\begin{tabular}{| c | c | c | c |c |}
\multicolumn{5}{c}{$b = 1$ $SU(N)$ matter possibilities}\\
 \hline
& & & &\\[-9pt]
$N$ & {\tiny\yng(1)}&${\tiny\yng(1,1)}$&
${\tiny\yng(1,1,1)}$ & {\rm neutral}\\[8pt]
\hline
$N \leq 24$ & $24 -N$ & 3 & 0 & $273-N (45-N)/2-1$\\
6 & $18 + k$ & $ 3-k $ & $k/2 , k \leq 3 $&$ 155-k $\\
7 & 22 & 0 & 1&132\\
\hline
\end{tabular}
\end{center}

We now show that global F-theory models can be realized for theories
with $SU(N)$ gauge group and all these possible matter representations
through the general construction described in the previous section,
except the special cases $N = 21, 23$.  Furthermore, the number of
neutral scalar fields in each of these models can be identified with
the number of unfixed parameters in the Weierstrass description of
each model when $N <  18$.

For $b =1$ on $\P^2$, the structure of the general Weierstrass model
is fairly simple.  Taking the locus of the $SU(N)$ to be the zero locus of
the function $\sigma = t$
 (in appropriate local coordinates $s, t$ on $\P^2$), the functions $f_i, i = 0,
\ldots, 12$ in the expansion of $f$ \eq{eq:fg-expansion} are
polynomials in $s$ of degree $12-i$, and the functions $g_i, i = 0,
\ldots, 18$ are polynomials in $s$ of degree $18-i$.  The functions
$f_i$ contain $1, \ldots, 13$ coefficients for a total of 91
coefficients while the $g_i$ contain 190 coefficients.  The total
number of coefficients appearing in the Weierstrass polynomials $f, g$
is therefore 281.  There is a redundancy in this description under
general linear transformations of homogeneous coordinates $s, t, u$ on
the F-theory base $\P^2$, removing 9 parameters.  The total number of
independent parameters in the Weierstrass model is therefore $281-9 =
272$.  There is one further scalar appearing in the low-energy 6D
theory associated with the overall K\"ahler modulus of the base, so
the number of scalar fields associated with the Weierstrass moduli is
in precise agreement with the gravitational anomaly condition, which states that
\begin{equation}
 H-V = 273 \,,
\label{eq:hv}
\end{equation}
where $H, V$ are the numbers of charged matter
hypermultiplets and vector multiplets in the generic model.

Now we apply the methods of Section~\ref{sec:global}.  Since $\{\sigma=0\}$
is a line in $\mathbb P^2$, all of the functions $\phi_0$, $\phi_1$,
\dots, etc.\ that occur in the analysis are in fact homogeneous polynomials
on $\mathbb P^2$ whose degrees are easily determined\footnote{For curves
of higher degree, particularly ones of higher genus, this statement
may fail to hold and the global analysis is more subtle.}.
Fixing the first few orders of the discriminant to vanish,
(\ref{eq:d0}) fixes the $13 + 19 = 32$ coefficients in $f_0, g_0$ in
terms of the four coefficients of $\phit$, thus removing 28
coefficients.  When the singularity locus is fixed at $t = 0$ this
removes two of the redundancies in the linear transformation
parameters.  Fixing $\Delta_1 = 0$ through (\ref{eq:d1}) removes
another 18 degrees of freedom by fixing $g_1$ in terms of $\phit,
f_1$, leaving $272-46 + 2 = 228$ degrees of freedom in the Weierstrass
coefficients\footnote{Note that the count we are performing here
only applies to $SU(N)$, $N\ge3$, since the Weierstrass coefficients
for $SU(2)$ involve $\phiorig$ rather than $\phit$.  In the case of
$SU(2)$, we use (\ref{eq:dorig}) to fix the $32$ coefficients in
$f_0, g_0$ in terms of the seven coefficients of $\phiorig$, removing
only 25 coefficients this time; (\ref{eq:d1}) still removes another
18 degrees of freedom, leaving $231$ degrees of freedom in the
Weierstrass coefficients.  The ``extra'' 3 degrees of freedom are
accounted for by the fact that the $\Lambda^2$ representation is trivial,
so the three copies of $\Lambda^2$ provide 3 additional neutral fields.
We thank Volker Braun for discussion on this point.}.
Fixing $\Delta_2 = 0$
through (\ref{eq:d2}) and (\ref{eq:d2bis}) removes another 20, bringing the number of
unfixed parameters in the $SU(3)$ model to 208.  This corresponds
precisely to the number of scalar fields (209) in the $N = 3$ model
from the table above.  Fixing $\Delta_3 = 0$ through (\ref{eq:d3})
removes another 19 parameters, leaving 189 degrees of freedom in the
Weierstrass coefficients, again in agreement with the 190 expected
scalar fields for the $SU(4)$
model above.
Note that the degrees of freedom in the Weierstrass coefficients are
complex degrees of freedom, while the hypermultiplets parameterize a
quaternionic K\"ahler moduli space and hence contain four real
scalars.  There are thus additional real degrees of freedom not
captured by the Weierstrass coefficients; these are associated with
degrees of freedom on the branes \cite{Bershadsky-jps}, and may be related to the T-brane
construction of \cite{T-branes}.

We now consider in more detail the matter content in
the set of theories with $SU(4)$ gauge group.  The 189
complex-dimensional moduli space of Weierstrass models with $SU(4)$
realized on a curve on $\P^2$ of degree $b = 1$ describes a family of
generic models with 3 matter fields in the two-index antisymmetric
($\Lambda^2$) representation.  This set of F-theory models satisfies
the conditions (\ref{eq:d0}-\ref{eq:d4}), and for a generic model in
this class there are three distinct roots of $\phit$ giving
singularity type
$4_b$.  For each such root, we can choose a local coordinate $s$ so
that $s = 0$ at the root, and we can expand
\begin{equation}
\phit = 2 s + {\cal O} (s^2) \,.
\label{eq:a3-phi-expansion}
\end{equation}
Plugging (\ref{eq:a3-phi-expansion}) into (\ref{eq:4-f}),
(\ref{eq:4-g}),
and
choosing $\phif = 0, f_2 = -1$ 
gives 
precisely the
expressions  (\ref{eq:a3-fg}) for $f, g$ used in
the $A_3 \rightarrow D_4$ singularity analysis of Section
\ref{sec:a3}.  For any $\phif, f_2$
an equivalent analysis will give a local singularity enhancement from $A_3$
to $D_4$ giving matter in the two-index antisymmetric ($\Lambda^2$)
representation.  Thus, these models all have 3 matter fields in the
$\Lambda^2$ representation, in agreement with the generic class of
models identified from the anomaly analysis.
The discriminant locus $\Delta$ is divisible by $\phit^4$, and the
remaining factor $\tilde{\Delta}_4$ is a degree 20 polynomial in $s$
and has 20 roots associated with singularities of type $4_a$ providing
20 fundamental representations, and completing the matter content of
these theories.  Though various non-generic singularities can be
constructed by tuning some roots of the discriminant to coincide, such
as the $E_6$ type singularity realized when $\phit = \phif = f_2 = 0$, the
anomaly analysis guarantees that such singularities cannot change the total matter
content of the theory as long as the gauge group remains $SU(4)$ and
no singularity becomes bad enough to provide an extra tensor
multiplet.

Continuing to higher $N$, the top class of models in the table above
is associated with generic singularities at the vanishing locus of
$\phit$, with no additional singularity structures.  As $N$ increases
up to $N = 17$, at each step an additional $3 + 20-N$ degrees of
freedom in the Weierstrass form are fixed, matching the decrease in
uncharged scalar degrees of freedom in the low-energy theory.
For small $N$ more restricted classes of Weierstrass coefficients
reproduce the other models in the table.

For $SU(5)$ the story is very similar to $SU(4)$.  There is a
171-dimensional space of models with 3 $\Lambda^2$ matter
hypermultiplets and 19 hypermultiplets in the fundamental
representation.

For $SU(6)$ the most generic model has $\phit = \alpha$, so there are
three singularities of type $6_b$ giving $\Lambda^2$ representations
and 18 fundamentals.  In this case, however, there are now other
possibilities.  Up to 3 of the roots of $\phit$ can be in $\beta$,
corresponding to singularities of type $6_c$, and giving
half-hypermultiplets in the $\Lambda^3$
representation.  This precisely reproduces the range of possible
$SU(6)$ models in the table above.  There are several interesting
features of these models.  First, consider the number of unfixed
Weierstrass degrees of freedom in these configurations.  From
\eq{eq:psf-ab}, \eq{eq:pf} we see that the number of degrees of
freedom in $\phis, \phif$ is reduced by 3 when fixing the $A_5$
singularity, independent of the distribution of roots between $\alpha$
and $\beta$.  From \eq{eq:f3e}, however, we see that the number of
degrees of freedom in $f_3$ ({\it i.e.}, in $\lambda$) is reduced by
one for each root of $\beta$.  Therefore, the dimension of the space
of models with $k$ roots $\beta = 0$ is reduced by $k$ from that of
the generic $SU(6)$ moduli space.  This agrees with the numbers of
neutral scalar fields listed in the table above for these models.
When $\beta = s$, as discussed in Section \ref{sec:a5}, the $E_6$
singularity is incompletely resolved, giving a half-hypermultiplet in
the $\Lambda^3$ representation.  When two roots of $\beta$ coincide,
however, we have $\beta = s^2$, giving a full $\Lambda^3$
hypermultiplet.

A further interesting feature of the $SU(6)$ models is the possibility
of a continuous phase transition between models with different numbers
of $\Lambda^3$ representations.  Consider a model with a (half-hypermultiplet) $\Lambda^3$
representation associated with a type $6_c$ singularity at a root $r$
of $\beta = 0$.  Such a model will also have an $A_6$ at every root of
$\phisa$.  By tuning one parameter, a root of $\phisa$ and the root
$r$ of $\beta$ can be made to coincide.  But at this point, this is a
common root of $\phit$ and $\phis$, and therefore from the definitions
of $\alpha$ and $\beta$ becomes a root of $\alpha$ and $\phifb$, and
also of $\lambda$, with $\alpha$ and $\lambda$ increasing in degree by
one and $\beta$ decreasing in degree by one.  At this point there is a
singularity of type $6_f$, as discussed in Section \ref{sec:global}.
From here, however, the
roots of $\alpha, \lambda$ and $\phifb$ can be freely and
independently varied.  This phase transition thus has the effect of
transforming matter between the representations
\begin{equation}
\frac{1}{2} \;{\tiny\yng(1,1,1)}+
{\tiny\yng(1)}\; \; \rightarrow \;
{\tiny\yng(1,1)} \,.
\end{equation}
This is not a simple Higgsing transition, since the gauge group does
not change.  There is no obstruction to such a transition from
anomalies, since the anomaly content of the matter representations is
the same on both sides of the transition.  We leave a further study of
this type of continuous F-theory transition between different types of
matter for further work.

For $SU(7)$, we again have a generic class of models of the correct
dimension with three $\Lambda^3$ representations.  There is also a
model of type 7B discussed in the previous section.  Since $\phit$ has
3 roots, in the decomposition \eq{eq:b-gd}, $\alpha, \beta, \gamma,
\delta$ can have respectively 1, 2, 0, 1 roots.  There is a single
singularity of type $7'_b$ in such models, associated with a single
$\Lambda^3$ representation.

We have thus reproduced all matter possibilities for $SU(N)$ models with $b =
1$.  We return to the discussion of the generic class of models for $N
\geq 8$.
As discussed above,
by tuning 3 parameters in an $f_i$ at each step through a relation like
\eq{eq:f4} or \eq{eq:s4}, and $20-N$ parameters through $g_{N -1}$, we
can continue to generate $A_{N -1}$ singularities up to a certain
point.  This continues up to $SU(17)$ without change, generating
models with these groups having three $\Lambda^3$ matter
representations and the correct number of degrees of freedom.  The
story changes slightly, however, at $SU(18)$.  At this point, the
equation analogous to \eq{eq:s4} would be $\psi_8 =-\frac13 \phit \phi_8$,
imposing the condition
that $\psi_8$ vanishes wherever $\phit$ vanishes.  Since $\psi_8$ is
linear, however, it must vanish.  But $\psi_8$ only has 2 degrees of
freedom, so the correspondence between the number of degrees of
freedom in the Weierstrass model and the number of neutral scalar
fields breaks down at this point.  We return to this point below;
nonetheless, we can continue to construct models with $SU(N)$ groups
beyond this point by setting $\psi_8 =
0$.  The next point where the analysis diverges from the general
pattern is at $SU(20)$.  At this point there is no further function
$g_{19}$ to fix, and $\psi_9$ is a scalar that we can set to 0.  This
is enough to guarantee vanishing of the discriminant to order 20.  At
the next order, fixing $f_{10}$ to match \eq{eq:expansion-f-even}
immediately guarantees vanishing to order 22, and fixing $f_{11}$ in
an analogous fashion gives a discriminant of order 24.  The
correspondence with the number of neutral scalar fields becomes quite
unclear  in these last steps, since in the 6D theories the number of neutral
scalars is expected to {\it increase} at 24 (with 20 neutral scalars
for $SU(24)$, 19 for $SU(23)$ and $SU(22)$, and 20 for $SU(21)$).

In any case, we can move directly to the end of the process just
described and write a general form for a class of models with $SU(24)$
local gauge group and three antisymmetric matter representations
\begin{eqnarray}
f & = &  -\frac13 \Phi^2 + \tilde{F}_{12}t^{12}\label{eq:exact-24}\\
g  & = &-\frac13\Phi f  -\frac1{27} \Phi^3 = \frac2{27}\Phi^3 -\frac13\Phi \tilde{F}_{12} t^{12}\nonumber\\
\Phi & = &  \left[\frac14 \phi_0^2 + \phi_1 t +\phi_2t^2+\phi_3t^3 
  \cdots +\phi_6t^6  \right]\,, \nonumber
\end{eqnarray}
where $\phi_0$ is a polynomial in $s$ of degree $3$,
 $\phi_k$ is a polynomial in $s$ of degree $6-k$ for $k>0$ and
$\tilde{F}_{12}$ is a constant (note that $G$ in \eq{eq:Tate-even} is
set to vanish in this class of models).
If $\tilde{F}_{12}$ is set to 0, then the model becomes everywhere
singular.

This is a good opportunity to comment on why our discussion has
always been about ``local'' gauge
groups.  The geometry of the singular fibers in an elliptic fibration
actually determines only the Lie algebra of the gauge theory, and there
are typically several different compact Lie groups with the same Lie
algebra.  (In the mathematics literature, these groups are said to be
``locally isomorphic.'')  The actual gauge group is determined by the torsion
in the Mordell--Weil group of the elliptic fibration \cite{pioneG}.
For the $SU(24)$ example just given, we show below that the
 Mordell--Weil group is in fact
the group $\mathbb Z_2$ with two elements, and this implies 
that the
true gauge group of the theory is $SU(24)/\mathbb Z_2$ rather than
$SU(24)$.

To see that this local $SU(24)$ example has a non-trivial Mordell--Weil group,
it is convenient to rewrite the example in Tate form, as in 
(\ref{eq:Tate-even}).  The result is
\begin{equation}
Y^2 + \phi_0 XY = X^3 + t \Upsilon X^2 + t^{12} \tilde{F}_{12}X  \,.
\end{equation}
where $\Upsilon = \phi_1  +\phi_2t+\phi_3t^2
  \cdots +\phi_6t^5 $.
The elliptic curve contains the point $(X,Y)=(0,0)$ and 
has a vertical tangent there,
for every value of $s$ and $t$. This implies by the
usual geometric law of addition on elliptic curves
\cite{silverman-tate}
that $(0,0)$ is a point of
order $2$ in the group law on each elliptic curve, so that the corresponding section defines
a point of order two in the Mordell--Weil group.

We have thus explicitly reproduced all the (local) $SU(N)$ models in the table
above, except $SU(23)$ and $SU(21)$.  
It is possible that
those two gauge groups can be realized through Higgsing of the
$SU(24)$ model or specialization of models with lower gauge groups.
It is also possible that the limitations we have encountered in
constructing F-theory models with these groups correspond to physical
constraints, perhaps associated with the discrete $\Z_2$ structure in
the $SU(24)$ theory.  Further analyses of these models, as
well as a precise understanding of the counting of degrees of freedom
for the space of models with large  $N$ are left for future work.

\vspace*{0.1in}

\subsection{$b = 2$}

We now consider the $T = 0$ 6D models with an $SU(N)$ gauge group and
$b = 2$.
For $b = 2$ the $SU(N)$ matter structures allowed by anomaly
cancellation are
\begin{center}
\begin{tabular}{| c | c | c | c |c |}
\multicolumn{5}{c}{$b = 2$ $SU(N)$ matter possibilities}\\
 \hline
& & & &\\[-9pt]
$N$ & {\tiny\yng(1)}&${\tiny\yng(1,1)}$&${\tiny\yng(1,1,1)}$
&${\tiny\yng(1,1,1,1)}$\\[12pt]
\hline
$N \leq  12$ & $ 48 -4N$ &  6 & 0& 0\\
6 & $24 + k$ &  $6-k$ & $k/2 \leq 3$& 0\\
7 &  20 + $5k$ &  $6-3k$ & $k\leq 2$& 0\\
8 & 25 & 2 & 1& 0\\
8 & $16 + 8k$ & $6-3k$ & 0 & $k/2 \le1$\\
\hline
\end{tabular}
\end{center}
In this case the analysis is slightly more complicated as we cannot
just take $\sigma = t$ and treat $f_i, g_i$ as functions of $s$, since
$\sigma$ is quadratic in $s, t$.  We
do not attempt to do a complete analysis constructing the most general
classes of models, but describe some simple salient features of the
models in this case.

The equation of a generic nonsingular degree two curve $\{\sigma=0\}$ can be put into the
form $\sigma = t^2 -s$ by choosing coordinates appropriately.  We can then do an expansion in $\sigma$ of
the form
$f =
f_0 + f_1 \sigma + \cdots$ where the $f_i$ are linear in $t$ and
otherwise generic polynomials in $s$.  Treating the expansions in this
way we can systematically carry out the analysis using the method
described in the previous section, since the ring of functions on
sufficiently small open subsets of
$\{\sigma=0\}$ is a UFD.  This becomes complicated in practice since at each
step we must use $t^2 \rightarrow s$ to bring products of functions
back to the canonical form where the coefficients in the $\sigma$
expansion are linear in $t$.   In principle, this approach leads to
constructions of general models with $b = 2$.

A non-generic class of such models is where we take $\sigma = t^2 -s$
with the $f_i$ being functions only of $s$.  This simplifies the analysis of
roots; the analysis is essentially as in the $b = 1$ case but each
function such as $\phit$ has twice as many roots when considered on
$\{\sigma=0\}$; for example, $\phit$ has six roots on  $\{\sigma=0\}$: $s = r, t = \pm
\sqrt{r}$ for each root $r$ of $\phit$ considered as a function of
$s$.  This leads to a construction of models precisely analogous to
those in the $b = 1$ case, including the 
models with six $\Lambda^2$ representations as well as the cases with
$\Lambda^3$ representations of $SU(6)$ and $SU(7)$.  Because this
simple class of models is not completely generic the number of
parameters is smaller than would be associated with the full moduli
space, and not all configurations are possible within this Ansatz.
In particular, because the roots of any function in $s$ are always
doubled in $\{\sigma=0\}$, we must get an even number of roots of $\beta$,
and the number of half-hypermultiplets for $SU(6)$ in the $\Lambda^3$
representation is always even.  Similarly, for $SU(7)$ the number of
$\Lambda^3$ representations is even, so we can get the model with 2
such representations but not the model with one.  To get the other
models with odd numbers of $SU(6)$ and $SU(7)$ $\Lambda^3$
representations it is necessary to go beyond this Ansatz.

A more generic class of $b = 2$ models can be identified following the
structure of \eq{eq:exact-24}.  We can construct a generic local $SU(12)$
model with six $\Lambda^3$ representations through
\begin{eqnarray}
f & = &  -\frac13 \Phi^2 + \tilde{F}_{6}\sigma^{6}\label{eq:exact-12}\\
g  & = &-\frac13\Phi f  -\frac1{27} \Phi^3 = \frac2{27}\Phi^3 -\frac13\Phi \tilde{F}_{6} \sigma^{6}\nonumber\\
\Phi & = &  \left[\frac14 \phi_0^2 + \phi_1 \sigma +\phi_2\sigma^2+\phi_3\sigma^3 \right]\,, \nonumber
\end{eqnarray}
where $\phi_i$ are in the ring of functions on $\{\sigma=0\}$.  
As in the $SU(24)$ case for $b=1$, putting the equation into Tate form
\begin{equation}
Y^2 + \phi_0 XY = X^3 + \sigma \Upsilon X^2 + \sigma^{6} \tilde{F}_{6}X  
\end{equation}
shows that $(0,0)$ is a point of order $2$ in the Mordell--Weil group, 
and hence the actual gauge
group is $SU(12)/\mathbb Z_2$.
Models with
smaller gauge groups can be found by adding higher order terms to $f,
g,$ to reduce the order of vanishing of $\Delta$.  By tuning
parameters in such models it should be possible to identify the $b =
2$ models with odd numbers of (half/full) $\Lambda^3$ hypermultiplets.

We have not identified the class of global F-theory
models giving rise to the
$\Lambda^4$ representation of $SU(8)$.  As discussed in Section
\ref{sec:local}, such matter representations should arise from a
singularity with a specific $A_7 \rightarrow E_8$ embedding.  Because
the $SU(8)$ model with a single $\Lambda^4$ representation (i.e., $k=2$
in the last line of the table above) does not contain
any $\Lambda^2$ representations, it seems that this model cannot arise
from a complete enhancement to $E_8$ through the embedding discussed
in Section \ref{sec:local}.  A related mechanism may be at work, however,
perhaps involving an incompletely resolved singularity.  We leave the
identification of the global $b = 2$ model with this matter structure
for further work.
We note, however, that since the $\Lambda^4$ representation of $SU(8)$
is quaternionic it can come in 1/2 hypermultiplet representations.  A
half hypermultiplet of $\Lambda^4$ combined with eight fundamental
representations has the same contribution to the anomalies as 3
$\Lambda^2$ representations.  We thus expect that there may be another
class of exotic transitions transforming matter in an $SU(8)$ gauge
group from
\begin{equation}
\frac{1}{2} \;{\tiny\yng(1,1,1,1)} \;+
8 \times \;{\tiny\yng(1)}\; \; \rightarrow \;
\; 3 \times \;{\tiny\yng(1,1)}\; \,.
\end{equation}

Finally, we identify another new type of phase transition associated with $b
= 2$ models.  Consider a class of $b = 2$ models with 
\begin{equation}
\sigma = t^2 -\epsilon s \,,
\end{equation}
where $\epsilon$ is a parameter for the models.  We can use the method
described above where each $f_i, g_i$ is a function purely of $s$ to
construct a subclass of the generic set of models with 6 $\Lambda^3$
representations of $SU(N)$.  Now we take the parameter $\epsilon
\rightarrow 0$.  This is just a parameter in the space of Weierstrass
models.  In the limit $\epsilon = 0$ this becomes a model with
a codimension one $A_{2 N -1}$  singularity localized on the zeros of
the function $\sigma' = t$.  This is
therefore identical to a $b = 1$ model with 3 $\Lambda^2$
representations of $SU(2N)$.  Considered in the opposite direction,
this transition provides a non-standard breaking of
an $SU(2N)$ theory with 3 $\Lambda^2$ representations to an $SU(N)$
theory with 6 $\Lambda^2$ representations.  A related transition has
recently been identified in the context of intersecting brane models
\cite{Nagaoka-Taylor}.  We leave a more complete discussion of this
type of phase transition for future work.

\vspace*{0.1in}

\subsection{$b =  3$}

For $b = 3$ the total genus (\ref{eq:genus-relation}) associated with the
matter content must be 1.  The only representations with genus 1 are
the adjoint and two-index symmetric (Sym${}^2$) representations.  So
each model must have one or the other of these.  We list the set of
possible matter contents for an $SU(N)$ theory with $b = 3$
\begin{center}
\begin{tabular}{| c | c | c | c |c |c |}
\multicolumn{6}{c}{$b = 3$ $SU(N)$ matter possibilities}\\
 \hline
& & & & &\\[-8pt]
$N$ &  ${\tiny\yng(1)}$ & ${\tiny\yng(1,1)}$ & ${\tiny\yng(1,1,1)}$ &Adj
& ${\tiny\yng(2)}$ \\[8pt]
\hline
$N \leq   8$ & $  72-9N$ &   9 & 0& 1 & 0\\
$N \leq   8$ & $  72-9N$ &   10 & 0& 0 &1\\
6 & $18 + k$ &  $9-k$ & $k/2 \leq 4$& 1 & 0\\
6 & $18 + k$ &  $10-k$ & $k/2 \leq 5$& 0 &1\\
7 & $9 +5k$ &  $9-3k$ & $k\leq 3$& 1 & 0\\
7 & $9 +5k$ &  $10-3k$ & $k\leq 2$& 0 &1\\
8 & 9 & 5 & 1& 1 & 0\\
8 & 9 & 6 & 1& 0 &1\\
9 & 5 & 4 & 1& 1 & 0\\
\hline
\end{tabular}
\end{center}
Note that the general pattern is that for $N > 5$, any number of
$\Lambda^3$ representations can be realized along with $(N -4) (N
-3)/2 -1$ extra fundamentals, at the cost of $N -4$ $\Lambda^2$
representations, beginning with the model with 9 $\Lambda^2$'s, one
adjoint, and $72-9 N$ fundamentals (or the same with 10 $\Lambda^2$'s
and one symmetric representation instead of the adjoint).  Such
exchanges are possible in the space of allowed theories except when
ruled out by the gravitational anomaly bound on scalar degrees of
freedom or positivity of the number of fundamentals; for example at
$SU(9)$ the number of fundamentals would become negative if we
attempted to remove the $\Lambda^3$ representation.  As in the $SU(6)$
case discussed above, we expect that all of these changes in matter
can be realized through phase transitions along continuous
one-parameter families of
F-theory models.

From the anomaly point of view, we can also exchange an adjoint
representation, along with one neutral scalar, for one symmetric and
one antisymmetric representation.  This cannot be done through
continuous phase transitions, however, since as discussed in Section
\ref{sec:local} the distinction between these representations is
determined by global monodromy on the brane structure.
Note that there are two models appearing
in the list of models with an adjoint that have no corresponding
model with  an Sym${}^2$ representation, the model with $SU(7), k = 3$ and that
with $SU(9)$.  In both cases this can be seen from counting degrees of
freedom.  These two models with the adjoint representation have a
total  of $273 + N^2 -1$ charged hypermultiplets.  Thus there are no
uncharged scalars in these models, by \eq{eq:hv}.  To exchange an adjoint for a
symmetric and an antisymmetric would require one additional charged
hypermultiplet, for a total of $273 + N^2$, violating the gravitational
anomaly bound.

As in the $b = 2$ case, we can proceed in several ways to construct
models of the generic $b = 3$ type with 9 $\Lambda^2$ representations
and one adjoint.  Choosing a generic cubic smooth $\sigma$, the
corresponding curve is an elliptic curve of genus one, giving one
adjoint representation.  We can expand order by order in the ring of
local functions on $\{\sigma=0\}$, or we can take a cubic such as $\sigma =t^3 +
s$ with non-generic coefficient functions depending only on $s$, or we
can construct the $N = 8$ model using an analogous construction to \eq{eq:exact-24},
\eq{eq:exact-12}.  By continuously deforming $\sigma$ we can get a
singular curve with an equation such as $\sigma = t^3 + st$ with a double point
singularity.  Because this is continuously connected to the family of
theories with smooth $\sigma$, however, this class of models should always have an
adjoint representation and not a symmetric representation.  We can
describe various models with $\Lambda^3$ matter content as discussed
in the $b = 2$ case above, though as in that discussion we cannot
explicitly identify all such models.  Note in particular that the
single $N = 9$ model cannot be realized in this way, and must require
some further tuning of the Weierstrass coefficients.  We leave a
further study of these models to future work.

\vspace*{0.1in}

\subsection{$b = 4$}

Now let us consider degree 4 curves, corresponding to $b = 4$ matter
content in the low-energy theory.  For $b = 4$, the total genus is 3.
So we expect 3 adjoints for a smooth degree 4 curve in F-theory.  From
the genus formula \eq{eq:genus}, the other possibilities for
saturating the genus are either a linear combination of $3-x$ adjoints
and $x$ Sym${}^2$ representations for arbitrary $SU(N)$, or several
exotic possibilities: a single ``box'' ( ${\tiny\yng(2,2)}$ )
representation for $SU(4)$ or a  ${\tiny\yng(2,1)}$  representation for
$SU(5)$; each have genus 3.

There are a variety of anomaly-free low-energy $SU(N)$ models with
various types of matter content, as in the cases with smaller $b$.
For $N \leq 6$ there are models with 3 adjoints, 12 $\Lambda^2$
representations, and no $\Lambda^3$ representations.  These correspond
to the generic branch in the Weierstrass models as described above and
can be constructed in a similar fashion to $b = 2, 3$.  There are a
variety of models that exchange $\Lambda^2$'s for $\Lambda^3$'s +
fundamentals.  We assume that these models correspond to various
singular limits in a similar fashion to that described above.  There
are also various models that replace some or all of the adjoints with
$\Lambda^2$ + ${\rm Sym}^2$ (again at the cost of a single neutral
scalar).  We do not have anything to say about these models that goes
beyond the discussion of the analogous models with $b = 3$.

The most novel feature that arises at $b = 4$ is the possibility of a
new matter representation as mentioned above.  Although there is no
apparently-consistent low-energy model that contains the
 ${\tiny\yng(2,1)}$  representation of $SU(5)$ at $b = 4$ (this
representation does appear for $SU(5)$ in combination with 3 adjoints
at $b = 5$), for $N = 4$
there is a model
\begin{equation}
SU(4): \;\;\;\;\; \;\;\;\;\;
{\rm matter} =
1 \times {\tiny\yng(2,2)}+
64 \times {\tiny\yng(1)}
\label{eq:4-box}
\end{equation}
While we identified a group-theoretic embedding of the box
representation of $SU(4)$ in Section \ref{sec:local}, we do not have
an explicit realization of a theory containing this representation as
a global Weierstrass model on $\P^2$.  Finding such a singularity may
involve an incomplete resolution of some kind, since the embedding
$A_3 \rightarrow D_6$ discussed in Section \ref{sec:local} would
otherwise seem to give rise to additional adjoint matter fields.  We
leave the construction of a global theory describing the model with
matter content \eq{eq:4-box} as a challenge for future work.

\section{4D models}
\label{sec:4D}

The general formalism developed for describing $SU(N)$ models in
Section \ref{sec:global} applies just as well to F-theory on an
elliptically fibered Calabi-Yau 4-fold as in the case of elliptically
fibered threefolds.  This provides a framework for systematically
analyzing F-theory constructions of 4D theories of supergravity
coupled to $SU(N)$ gauge theories.  For 4D F-theory constructions the
full story is more complicated, since fluxes must be present
\cite{Becker-m, Dasgupta-rs}.   The fluxes
generate a superpotential, and nonperturbative contributions from
instantons are also present.  These effects produce a potential on the
moduli space that lifts the continuous flat moduli space to a
landscape with separated vacua and stabilized moduli.  Nonetheless,
underlying this more complicated physics is the continuous moduli
space of degrees of freedom associated with the Weierstrass
coefficients in an F-theory construction.  When the compactification
space is large, these moduli will be light, and the moduli space is
approximate.

\subsection{4D Weierstrass models}
\label{sec:4D-Weierstrass}

We do not go far into the issues regarding moduli stabilization and
fluxes on 4D F-theory vacua here.  F-theory
methods for analyzing matter in 4D
theories in the presence of flux were developed 
in \cite{Donagi-Wijnholt,Beasley-hv}; following these works
there has been a great deal of recent work on 4D F-theory
constructions with particular focus on phenomenological applications
(see for example \cite{Beasley-hv2, Marsano-F-theory,
  Blumenhagen-F-theory, Cvetic-gh}); for reviews of some recent developments in
these directions see \cite{Denef-F-theory, Heckman-review,
  Weigand-review}.  In this paper we take a simplistic approach where
we ignore fluxes and the lifting of moduli, and consider the tuning
necessary in Weierstrass models to achieve an $SU(N)$ gauge group.  We
can then consider constructions with matter fields in different
representations.  Although the number of fields appearing in a
particular representation may depend upon the details of fluxes and
the full F-theory construction, the type of representation should
depend only on the classification of codimension two singularities, on
which we are focused here.

In four dimensions, as in six dimensions, the simplest F-theory
compactification we may consider is compactification on projective
space.  We thus consider F-theory on a 4-fold that is elliptically
fibered over $\P^3$.  We consider some explicit examples of the
structure of Weierstrass models giving ${\cal N} = 1$ 4D supergravity
theories in this context.  Previous work in which F-theory
constructions over $\P^3$ were considered includes \cite{Klemm-lry}.

On $B =\P^3$ we have $K = -4H$, where $H$ is the hypersurface divisor
generating $H_2 (B,\Z)$.  The Weierstrass functions $f, g$ are then
polynomials of degree $16$ and 24 in local variables $r, s, t$, and
the discriminant is of degree 48.
We are looking for an $SU(N)$ gauge group associated with an $A_{N
  -1}$ singularity.  We consider the discriminant locus on a degree
$b$ hypersurface $\{\sigma=0\}$.  The coefficient functions
$f_i (g_i)$ then have degree $16-bi, 24 -bi$

We begin as in 6D with $b = 1$.  We again follow the systematic
analysis of Section \ref{sec:global}, using $\sigma = t$, so that $f,
g$ are functions of $r, s$.  The function $\phit$ controlling the
leading term $f_0$ is now of degree 4.  We can construct generic models with
$SU(N)$ gauge groups by tuning the coefficients to make each term
$\Delta_n$ in the discriminant vanish order by order, as in Section
\ref{sec:global}.  In the generic model, matter will be associated
with the points where $\Delta_{N}$ acquires extra degrees of
vanishing, associated with codimension two singularities.  The
intersection between $\{\sigma=0\}$ and $\{\phit=0\}$ defines a curve in $\P^3$
that is generically a genus 3 curve.  There will be matter in the
2-index antisymmetric $\Lambda^2$ representation of $SU(N)$ localized
on this curve.  As mentioned above, a precise determination of the
number of matter fields in this representation depends on details of
the theory such as fluxes that we do not consider here.  The rest of $\Delta$
defines another divisor (possibly reducible) whose intersection with
$\{\sigma=0\}$ gives another curve (possibly disconnected) that supports
matter in the fundamental representation.  Although the curve $\{\phit=
 \sigma=0\}$ is of higher genus, at generic points along this curve
the singularity is a codimension two singularity identical to the
$A_{n -1}\rightarrow D_{n}$ singularities discussed earlier.

The generic 4D model with $b = 1$ having largest gauge group can be
described in a fashion similar to \eq{eq:exact-24}
\begin{eqnarray}
f & = &  -\frac13 \Phi^2 + \tilde{F}_{16}t^{16}\label{eq:exact-32}\\
g  & = &-\frac13\Phi f  -\frac1{27} \Phi^3 = \frac2{27}\Phi^3 -\frac13\Phi \tilde{F}_{16} t^{16}\nonumber\\
\Phi & = &  \left[\frac14 \phi_0^2 + \phi_1 t +\phi_2t^2+\phi_3t^3 
  \cdots +\phi_8t^8  \right]\,, \nonumber
\end{eqnarray}
where $\phi_0$ is a polynomial in $r, s$ of degree $0$,
 $\phi_k$ is a polynomial in $r, s$ of degree $8-k$ for $k>0$ and
$\tilde{F}_{16}$ is a constant.
Once again, we can write this in Tate form
\begin{equation}
Y^2 + \phi_0 XY = X^3 + t \Upsilon X^2 + t^{16} \tilde{F}_{16}X  \,.
\end{equation}
to see that $(0,0)$ is a point of order $2$ in the Mordell--Weil group
of the elliptic fibration.  Thus, the gauge
group in this case is $SU(32)/\mathbb Z_2$. 
 The curve supporting the $\Lambda^2$
matter is the intersection between $\{\sigma=0\}$ and $\{\phi_0=0\}$.  Models with
smaller gauge group can be found by adding high-order terms to $f, g$
to reduce the order of vanishing of the discriminant $\Delta$.

Just as in 6D, the parameters of the theory can be tuned so that there
are more elaborate codimension two singularities in the 4D $SU(N)$
models.  For an $SU(6)$ model, for example, as in \eq{eq:pt-ab}, if
$\phit$ does not divide $\psi_2$, then there must be a component
$\beta$ of $\phit$ that is a factor of $\phi_1$.  The intersection of
$\{\beta = 0\}$
with $\{\sigma=0\}$ gives a curve supporting matter in the $\Lambda^3$
representation of $SU(N)$.  Since $\{\sigma=0\}$ is smooth, and $\Delta$ is
smooth at generic points, for $b = 1$ the only general classes of
codimension two singularity types are the same as those that can
arise for $b = 1$ models in 4D, namely $n$-index antisymmetric matter
fields.  As we discuss further in the following subsection, this gives
a constraint (though relatively mild) on certain classes of 4D ${\cal
  N} = 1$ supergravity theories that can be realized in F-theory.

For higher $b$, the story is again parallel to that in 6D, although
our understanding of the details such as the number of types of each
matter field is not as complete without a careful treatment of
fluxes.  Nonetheless, just as in 6D, matter with a nonzero genus
contribution $g_R$ can only arise when $b > 2$, and will be associated
with codimension one singularities on $\{\sigma=0\}$.

\subsection{A (mild) constraint on 4D supergravity theories}

The above analysis leads to a constraint on the set of 4D ${\cal N} = 1$
supergravity theories that can be realized from F-theory.  This
constraint is rather specific to the models associated with the $\P^3$
compactification, but serves as an example of a constraint on possible
low-energy 4D supergravity models.

From the point of view of the 4D theory, the constraint is of the form
``any theory with property $X$ has features $Y$,'' where $X$ describes
a set of
properties that uniquely determine the F-theory construction to come
from an elliptic fibration over $\P^3$ with a gauge group $SU(N)$
realized on a divisor $\{\sigma=0\}$ of degree $b = 1$, and $Y$ are the
constraints on models of this type.

We briefly summarize the features ($X$) of a 4D model that uniquely
determine the F-theory base and $SU(N)$ divisor class to be $\P^3$
and $\{\sigma=0\} = H$.  

We begin with the correspondence between discrete structures in the 4D
supergravity theory and in the base of the F-theory compactification;
the connection between the F-theory geometry and the low-energy theory
is
systematically described in \cite{Grimm-F-theory}, and
further analysis of this correspondence will appear in
\cite{Grimm-Taylor}.  Similar to the story in 6D, a 4D F-theory
compactification on a base $B$ gives rise to topological terms in the
low-energy action of the form
\begin{eqnarray}
\tau_R &  \sim &  -K \cdot\chi \; \tr R \wedge R   \label{eq:topological-couplings}\\
\tau_F &  \sim &  b \cdot\chi \;\tr F \wedge F \,, \nonumber
\end{eqnarray}
where $\chi$ are axions coming from wrapping the $C_4$ Ramond-Ramond
field on divisors of the base.  In 6D, the corresponding terms appear
with two-form fields in place of axions, since $C_4$ is wrapped on
2-cycles instead of 4-cycles.  The $\tau_F$ term is simply the usual
coupling between $C_4$ and 3-branes associated with instantons on the
7-branes, where $b$ is the divisor class of the 7-branes carrying the
local factor of the gauge group.  The $\tau_R$ term comes from the
coupling of the 7-branes to curvature, summed over all 7-branes as
described in the 6D case by Sadov \cite{Sadov}, and $K$ is the
canonical class of the base.  The number of axions of this type is
given by the Hodge number $h^{(1, 1)} (B)$ of the F-theory base.  For
$\P^3$, $h^{1, 1} = 1$ and there is only one such axion.  In general,
$K, b$ are elements of a lattice $L$, where the shift symmetries of
the axions live in the dual lattice $L^*$.  In the case of only one
axion $\chi$ such as for F-theory on $\P^3$, the couplings in $\tau_R,
\tau_F$ are each quantized so that $K, b$ are integers.

Now let us consider the special features of the base $\P^3$ that may
be visible in the 4D supergravity theory.  There are a number of
spaces with $h^{1, 1} = 1$ that could act as bases for a 4D F-theory
compactification.  Any such space must be Fano, since $-K$ must be
effective (though note that F-theory bases with $h^{1, 1} > 1$ need
not be Fano).  Fano spaces with $h^{1, 1} = 1$ have been completely
classified \cite{Iskovskih}.  For such a space, the {\sl index} is the ratio
$-K/x$ where $x$ is the smallest effective divisor class, the
generator of $H^2 (B,\Z)$.  For projective space $\P^3$, the index is
4, since $K = -4H$.  All other Fano spaces with $h^{1, 1} = 1$ have a
smaller value of the index.  The ratio between the integers
parameterizing the topological couplings \eq{eq:topological-couplings}
is the ratio $-K/b$ between the canonical class of the base and the
divisor class characterizing each local factor of the gauge group.  
This ratio must
be less than the index of the F-theory base, since $b \geq 1$.  Thus,
for theories with $\ho = 1$, the maximum value of $-K/b$ possible is
4, and this value is only attained when the base is $\P^3$ and the
local factor of the gauge group 
is wrapped on the divisor $H$, corresponding to the
case $b = 1$ analyzed above.

Thus, we can state a weak constraint on 4D supergravity theories that
come from F-theory: any 4D ${\cal N} = 1$ supergravity theory with
only one of the appropriate type of axion, and couplings
\eq{eq:topological-couplings} with a ratio of integers $-K/b = 4$
that has an $SU(N)$ local gauge group  factor must have $N \leq 32$, and can
only have matter in $k$-index antisymmetric representations of
$SU(N)$.  In particular, such a theory cannot have matter in the
adjoint representation of the gauge group.

This is not a strong constraint.  And there are a number of rather
subtle issues in  making this constraint rigorous.  In
particular, the lifting of the moduli by the flux and nonperturbative
superpotential make the determination of the spectrum and terms in the
action less clear than in 6D theories where the spectrum must be
massless.  Nonetheless, at least for large volume compactifications
the structure of the theory determined by F-theory should be apparent
in the low-energy theory, and at least in this regime this constraint
should hold.  

Despite the limitations in the range of applicability and
interpretation of this constraint, it is  interesting to study the
constraints that
F-theory places  on 4D supergravity theories.  It
should not be surprising that such constraints exist; string constructions generally place many
constraints on which possible low-energy theories can be realized.  In
six dimensions, anomalies provide a window on the strong constraints
imposed by F-theory constructions  \cite{universality}-\cite{0}, and other F-theory
constraints can also be identified as consistency conditions from the
point of view of the low-energy theory \cite{Seiberg-Taylor}.  Further
discussion of constraints on 4D theories from F-theory will appear in
\cite{Grimm-Taylor}.  It will be interesting to investigate whether the
type of constraint on gauge group and matter content identified in
this paper can be generalized and understood in terms of macroscopic
consistency conditions from the point of view of 4D supergravity.

\section{Conclusions}
\label{sec:conclusions}

We have explored the structure of some codimension two singularities
in F-theory and the matter representations to which they give rise.
The focus here has been on understanding how such codimension two
singularities arise in global F-theory models.  We have developed a
very general characterization of global Weierstrass models giving rise
to $SU(N)$ gauge groups, and analyzed how this general framework
applies for F-theory constructions on the bases $\P^2$ and $\P^3$.

It is clear that there is still much unexplored territory in the full
range of codimension two singularities.  Beyond the standard rank one
enhancement studied by Katz and Vafa, there are singularities with
incomplete resolution, higher rank enhancement, and singularities
associated with singular curves in the base, all of which can give
rise to different kinds of matter in F-theory constructions.  Further
exploring this range of possibilities should provide a fruitful
enterprise for further understanding the r\^ole of matter in F-theory
and string theory.

One interesting feature that we have encountered here is the presence
of novel phase transitions in F-theory.  We have identified
phase transitions in which a matter field transforming in the
3-index antisymmetric representation of $SU(6)$ combines with a matter
field in the fundamental representation
to produce a matter field in the 2-index
antisymmetric representation.  This transition does not change the
gauge group and hence is not a standard Higgsing transition,  but should
have some description in the low-energy field theory.  There are
analogous transitions for the 3-index antisymmetric representation of
any $SU(N), N \geq 6$.
We expect similar transitions for other recombinations of matter fields
that leave the 6D anomaly contributions unchanged, such as transitions
involving the
4-index representation of $SU(N), N \geq 8$.  We have also
found unusual transitions where the group $SU(2N)$ breaks to $SU(N)$
with three matter fields in the two-index antisymmetric representation
going to six such fields in the $SU(N)$ theory.  We hope to return to
a more detailed study of these exotic phase transitions in future
work.

Using global F-theory models on the base $\P^2$  to describe 6D
supergravity theories without tensor multiplets, we have shown that a
systematic parameterization of Weierstrass models precisely matches
the space of theories identified through anomaly constraints in the
low-energy theory, at least for $SU(N)$ gauge groups supported on
curves of low degree in the F-theory base.  The structure of matter
representations in these theories and number of degrees of freedom
matches neatly between F-theory and the low-energy analysis for small
$N$ and degree, with more complicated phenomena arising at higher $N$
and degree that pose interesting questions for future work.

Applying the global analysis of Weierstrass models to 4D F-theory
constructions we have characterized the matter content of a simple
class of $SU(N)$ models on $\P^3$.  This leads to a mild constraint on
4D supergravity theories, limiting the gauge group and matter content
for this specific class of models.
This class of models can be identified from the spectrum and
topological couplings of the 4D theory.  Further work in this
direction promises to expand our understanding of F-theory constraints
on 4D supergravity theories, and to clarify the structure of matter
fields in general F-theory constructions.
\vspace*{0.3in}

\noindent
{\bf \large Appendix}

\appendix
\section{Details of singularity resolutions}

In this appendix we give detailed analyses of the singularity
resolution of various codimension two singularities described in the
main text.  We proceed by considering the blow-ups in a sequence of
local charts.  Note that in these analyses we choose a minimal set of
charts to resolve the singularities.  In analyzing any given
situation, it is generally necessary to check all charts for
additional singularities to be sure of a complete resolution.

\subsection{Enhancement of $A_3$ on a smooth divisor class}
\label{sec:a3-appendix}

As a first example we consider the Weierstrass model for the
codimension two singularity enhancement $A_3 \rightarrow D_4$.  
As discussed in the main text, after a change of variables a
particular form of the local
Weierstrass equation for a singularity enhancement of this type is
\begin{equation}
\Phi =
-y^2 + x^3 + s^2 x^2 -t^2 x = 0 \,.
\label{eq:a3-appendix}
\end{equation}
This is a local equation for the Calabi-Yau threefold described by an
elliptic fibration where $s, t$ are local coordinates on the base $B$.
At $t = 0$ there is a codimension one $A_3$ type singularity that
becomes a $D_4$ singularity on the $s = 0$ slice.

\subsubsection{Resolution of $A_3$}
\label{sec:a3-resolution}

The threefold given by (\ref{eq:a3-appendix}) is singular along the locus $t =
0$ at $x = y = 0$ for all values of $s$.  This singularity can be
resolved using a standard procedure of blowing up the codimension one
singularity repeatedly until the space is smooth.  We do this by
working in a sequence of charts containing the various blow-ups.  We
go through this process in  detail in this case, as all other
examples will follow in a similar fashion.  In this part of the
analysis we fix $s \neq 0$.  We are thus essentially working on a surface
that is a two complex dimensional slice of the full Calabi-Yau
threefold.
\vspace*{0.05in}

\noindent
{\bf Chart 0}: 

In the original chart we have coordinates $(x, y, t)$ (treating $s$ as
a constant), and the equation (\ref{eq:a3-appendix}) gives a singularity at $x
= y = t = 0$.
\vspace*{0.05in}

\noindent
{\bf Chart 1}: 

To resolve the singularity in chart 0, we blow up the singular point,
replacing
the point $(0, 0, 0)$  with a $\P^2$ given by the set of limit points
described by homogeneous coordinates
$[x, y, t]$ along curves approaching $(0, 0, 0)$.  We choose a local
chart that includes the points $[x, y, 1]$ by changing coordinates to
\begin{equation}
(x, y, t) = (x_1 t_1, y_1 t_1, t_1) \,.
\end{equation}
In these coordinates, the local equation (\ref{eq:a3-appendix})
becomes
\begin{equation}
\Phi =
(-y_1^2 + s^2 x_1^2 + x_1^3t_1-t_1x_1) t_1^2 = 0\,.
\label{eq:a3-t1}
\end{equation}
The $\P^2$ that is added through the blow-up process is known as the
{\it exceptional divisor} associated with the blow-up.  Factoring out
the overall $t_1^2$ from (\ref{eq:a3-t1}) ({\it i.e.}, removing two
copies of the exceptional divisor), we have the equation for the
{\it proper transform} of the original space
\begin{equation}
\Phi_t =
-y_1^2 + s^2 x_1^2 + x_1^3t_1-t_1x_1  = 0\,.
\label{eq:a3-t}
\end{equation}
This equation describes the Calabi-Yau space in chart 1 after the
original singularity at $(x, y, t) = (0, 0, 0)$ has been blown up.
(We use the subscript to denote the coordinate chart used for the blow-up.)
The intersection of the space defined through (\ref{eq:a3-t}) with the
exceptional divisor at $t_1 = 0$ gives the exceptional divisor on the
Calabi-Yau threefold, which (on the surface associated with the slice
at fixed $s$) is generally a curve or set of curves
associated with blowing up the point at $t_1 = 0$.  At $t_1 = 0$,
(\ref{eq:a3-t}) becomes $-y_1^2 + s^2 x_1^2 = 0$ or
\begin{equation}
y_1 = \pm s x_1 \,.
\label{eq:a3-c}
\end{equation}
This defines a pair of curves that we call $C_1^{\pm}$.
The equation (\ref{eq:a3-t}) still contains a singularity at the point
$(x_1, y_1, t_1)= (0, 0, 0)$, where the curves $C_1^\pm$ cross.  So we
must again blow up the singularity to produce a smooth space.
\vspace*{0.05in}

\noindent
{\bf Chart 2}: 

We replace the singular point in Chart 1 with another exceptional divisor
$\P^2$, this time using the local coordinates
\begin{equation}
(x_1, y_1, t_1) = (x_2, y_2  x_2, t_2 x_2)  = 0\,.
\end{equation}
After removing two copies of the exceptional divisor $x_2 = 0$
we get the new
local equation
\begin{equation}
\Phi_{tx} =
-y_2^2 + s^2 + x_2^2 t_2-t_2  = 0\,.
\label{eq:a3-tx}
\end{equation}
This gives another exceptional curve $C_2$ (on the surface at each
$s$), associated with the intersection of (\ref{eq:a3-tx}) with the
exceptional divisor $x_2 = 0$
\begin{equation}
C_2 =\{(x_2, y_2, t_2): x = 0,  t_2 = s^2 -y_2^2\} \,.
\label{eq:a3-c2}
\end{equation}
In homogeneous coordinates on $\P^2$, $C_2$ is given by the set of
points $[1, y_2, s^2 -y_2^2]$
Since (\ref{eq:a3-tx}) has no further singularities, we have
completely resolved the local singularity and have a smooth space in
coordinate chart 2.

From the way in which the exceptional curves $C_1^\pm, C_2$ intersect,
we identify the $A_3$ form of the singularity found by Kodaira.  To
compute the intersections, we write the equation (\ref{eq:a3-c}) for
$C_1^\pm$ in terms of coordinates in chart 2
\begin{equation}
y_2x_2 = \pm sx_2 \; \Rightarrow
\; y_2 = \pm s \,,
\end{equation}
which combined with $t_1 = t_2x_2 = 0$ gives the points $[1, \pm s,
  0]$ in homogeneous coordinates on the $\P^2$ containing $C_2$,
showing that $C_1^\pm$ each intersect $C_2$ at a single point but do
not intersect one another, corresponding to the structure of the $A_3$
Dynkin diagram.

\subsubsection{Resolution of local $D_4$ singularity on $s = 0$ slice}

We now return to the form (\ref{eq:a3-appendix}) for the elliptic fibration
with a $D_4$ singularity at the point $s = 0$.

We begin by confirming that on the slice $s = 0$ there is indeed a
singularity whose resolution is described by a set of curves with
$D_4$ structure.  To see this  we must resolve the singularity given by
\begin{equation}
\Phi =
-y^2 + x^3 -t^2 x = 0 \,.
\label{eq:d4-appendix}
\end{equation}
Following essentially the same procedure as in the $A_3$ case, we
blow up the singularity at $(x, y, t) = (0, 0, 0)$ by passing to a
chart 1 with
\begin{equation}
(x, y, t) = (xt, yt, t)_1 \,.
\end{equation}
Here and in the following examples we will streamline notation by
not explicitly including the subscripts in each chart except when
necessary.  In chart 1, the equation becomes
\begin{equation}
\Phi_t =
-y^2 + x^3t-tx = 0 \,.
\end{equation}
The exceptional divisor $\delta_1$ at $t = 0$ is then given by $y^2 =
0$ so
\begin{equation}
\delta_1 =\{(x, 0, 0)\} \,.
\label{eq:d1-exceptional}
\end{equation}
There are still singularities at $t = 0$ when $x^3-x = 0$, so at the points
\begin{equation}
x = 0, \pm 1 \,.
\end{equation}
Blowing up each of these three singularities gives three further curves
$\delta_2^{0, \pm}$, each of which intersects with $\delta_1$, and
which do not intersect with each other, giving the familiar $D_4$
singularity resolution.

\subsubsection{Enhancement $A_3 \subset D_4$}

Now, let us go through this analysis more carefully for the full
threefold incorporating the coordinate $s$.  This will enable us to
understand how the $A_3$ structure is embedded in the $D_4$
exceptional curves, giving an explicit characterization of the
resulting matter structure in terms of group theory.
We wish then to resolve all singularities in the Calabi-Yau threefold
defined by
\begin{equation}
\Phi =
-y^2 + x^3 + s^2 x^2 -t^2 x = 0 \,,
\label{eq:a3b}
\end{equation}
including $s$ as a coordinate in the analysis.
\vspace*{0.05in}

\noindent
{\bf Chart 1}:

At the first stage in analysis, the coordinate $s$ can be carried
along as a spectator variable in passing from chart 0 to chart 1,
since the point $(x, y, t) = (0, 0, 0)$ is singular for all $s$.
Thus, we use the coordinate change
\begin{equation}
(x, y, t, s) = (xt, yt, t, s)_1 \,.
\label{eq:blow-up-t1}
\end{equation}
In chart 1 the full equation then becomes
\begin{equation}
\Phi_t =
-y^2 +  s^2 x^2 +x^3t-tx = 0 \,.
\label{eq:a3-c1}
\end{equation}
We see from this that the exceptional curves $C_1^\pm$ defined by $y =
\pm sx$
(\ref{eq:a3-c}) indeed collapse at $s = 0$ to the same curve
$\delta_1$ given by $y = 0$ (\ref{eq:d1-exceptional}).
Singularities arise at $t = 0$ in (\ref{eq:a3-c1}) for all $s$ at $x =
0$, and for $s = 0$ at $x = \pm 1$.  
\vspace*{0.05in}

\noindent
{\bf Chart $2_0$}: 

The singularity at $ x = 0$ can
again be handled by blowing up at each $s$, going to coordinate chart
$2_0$ given by
\begin{equation}
(x, y, t, s)_1 = (x, y x, t x, s)_2 \,,
\label{eq:blow-up-x1}
\end{equation}
where
\begin{equation}
\Phi_{tx} = -y^2 + s^2 + x^2 t-t = 0 \,.
\end{equation}
In this chart there is a single new exceptional curve given by
$x = 0,  t = s^2 -y^2$, which for generic $s$ is the curve $C_2$ from
(\ref{eq:a3-c2}), and which for $s = 0$ is the curve $\delta_2^0$
given by $x = 0, t = y^2$.
\vspace*{0.05in}

\noindent
{\bf Conifold-type double point singularities at $x = \pm 1$}:

The story is a little more interesting at the singular points $x = \pm
1, s = 0$ of (\ref{eq:a3-c1}).  For example, shifting the coordinate
$x \rightarrow x +1$ to place the $ x =1$ singular point at the
origin, (\ref{eq:a3-c1}) becomes
\begin{equation}
-y^2 + s^2 (x +1)^2 + t x (x +1) (x + 2) = 0\,.
\end{equation}
Near the singular point $(x, y, t, s) = (0, 0, 0, 0)$ this singularity
has the form
\begin{equation}
(s-y) (s + y) + 2tx = 0.
\end{equation}
This is the familiar ordinary double point singularity that appears on
conifolds \cite{conifold}, and that has
played a fundamental r\^ole in understanding many aspects of string
theory vacua.  This singularity can be resolved in two different ways
to locally give a smooth Calabi-Yau threefold.  The resolution can be
done by replacing the singular point with a curve $\P^1$ either
by blowing up $t
= 0, s-y = 0$ or by blowing up $t = 0, s + y = 0$.
In either case, we get an additional exceptional curve at $s = 0$
that we can call $\delta_2^+$.  
Now, let us consider how the curves
$C_1^\pm$ relate to $\delta_2^+$.  After the coordinate shift
$x\rightarrow x +1$, $C_1^\pm$ are given by
\begin{equation}
y = \pm s (x +1) \,.
\end{equation}
If we resolve the local singularity with a curve by blowing up $t =s-y = 0$, then
homogeneous coordinates on the new $\P^1$ are given by $[x, s + y]$.
In the limit $x, y, t, s \rightarrow 0$,
the points $[x, 2s + sx]\sim[x, 2s]$ will be included for $C_1^+$, 
while the points $[x, -xs]\sim[1, 0]$ will be included for $C_1^-$.  This shows
that in this case, the proper transform of $C_1^+$ contains all of
$\delta_2^+$, while
the proper transform of $C_1^-$ only contains one point in
$\delta_2^+$.  If we make the other choice for blowing up the conifold
singularity, then $C_1^-$ contains $\delta_2^+$ while $C_1^+$ does
not.  A similar analysis holds for the resolution of the singularity
at $x =-1$ in (\ref{eq:a3-c1}).

The results of this analysis can be summarized as follows: there are
two choices that can be made in blowing up each of the two
conifold singularities in
the full Calabi-Yau threefold given by the local form of the elliptic
fibration (\ref{eq:a3-appendix}).  Parameterizing these choices by $\tau_+,
\tau_-\in\{0,1\}$ and denoting $\bar{\tau} = 1-\tau$, we have an
explicit embedding of $A_3$ into $D_4$ through
\begin{eqnarray}
C_1^+ & \rightarrow &  \delta_1 + \tau_+ \delta_2^+ + \tau_-\delta_2^- \nonumber\\
C_1^- & \rightarrow &  \delta_1 + \bar{\tau}_+ \delta_2^+ + 
\bar{\tau}_-\delta_2^-
\label{eq:a3d4}
\\
C_2 & \rightarrow &  \delta_2^0 \,. \nonumber
\end{eqnarray}
It is straightforward to check that for any of the four possible
choices of combinations of $\tau_\pm$, the intersection form of $A_3$
is correctly reproduced by this embedding.  For example, for $\tau_+ =
\tau_-= 1$
\begin{equation}
C_1^+ \cdot C_1^-= (\delta_1 + \delta_2^+ + \delta_2^-)\cdot \delta_1
= 0 \,,
\end{equation}
using the fact that $\delta_1 \cdot \delta_1 = -2$ since $\delta_1$ is
a genus 0 curve.
The embedding (\ref{eq:a3d4}) for choice of parameters $\tau_+ =
\bar{\tau}_-= 1$ is depicted in Figure~\ref{f:a3d4}.

From this embedding of $A_3$ into $D_4$ we can read off the matter
content in terms of the representation theory of $SU(4)$.  Each new
genus 0 curve that is added to the threefold at the point $s = 0$
that does not appear at generic $s$ represents a matter field whose
transformation under the $A_3$ gauge group is determined by the
intersection form with the curves forming the $A_3$ structure.  This
is the F-theory version \cite{Katz-Vafa} of the way in which shrinking
2-cycles produce charged matter in type II \cite{enhanced} and M-theory \cite{Witten-mf}.  In
this case, the curves $\delta_1, \delta_2^{0, \pm}$ form a basis of
simple  roots for the $D_4$ algebra.  Thus, the complete set of genus
0 curves at $s = 0$ corresponds to the set of all roots in $D_4$.
Since the embedding $A_3 \subset D_4$ is unique up to isomorphism,
this corresponds to the standard Katz-Vafa picture in which the gauge group is
enhanced by rank 1, and the matter fields are given by the weights of
the adjoint representation of $D_4$ as they transform under $A_3$
(leaving out the adjoint of $A_3$, which corresponds to the generators
of the $SU(4)$ gauge group itself).
As discussed in the main text, this gives
a matter field in the  two-index antisymmetric
representation ( ${\tiny\yng(1,1)}$ , or $\Lambda^2$) of $SU(4)$.

\subsection{Enhancement of a local $A_5$ singularity}
\label{sec:a5-appendix}

Now, we consider the enhancement of $A_5$ by various types of local
singularities and the associated matter content.  

\subsubsection{Enhancement $A_5 \subset D_6$}
\label{sec:a5-d6}

We begin with the local equation \eq{eq:phi-a5-d6}
\begin{equation}
\Phi = -y^2 + x^3 +s^2 x^2 +3x^2 t +
 3t^3x + 2s^2 t^2 x + s^2 t^4  = 0\,.
\label{eq:phi-a5-d6-appendix}
\end{equation}
\vspace*{0.05in}

\noindent
{\bf Chart 1}: 

Blowing up in the $t$ chart (\ref{eq:blow-up-t1}) gives
\begin{equation}
\Phi_t = -y^2 + x^3 t +s^2 x^2 +3x^2 t +
 3t^2x + 2s^2 t x + s^2 t^2 = 0\,.
\label{eq:a5-1}
\end{equation}
The exceptional divisor at $t = 0$ on the threefold  is
\begin{equation}
C_1^\pm =\{(x, \pm sx, 0, s)\}
\end{equation}
for generic $s$, which degenerates to
\begin{equation}
\delta_1 =\{(x, 0, 0, 0)\}
\end{equation}
at $s = 0$.
There are singularities in (\ref{eq:a5-1}) at $(0, 0, 0, s)$ for all
$s$ and an additional singularity at $(-3, 0, 0, 0)$ at $s = 0$.

Dealing with the isolated singularity at $s = 0, x = -3$ first, we
change coordinates $x \rightarrow x + 3$, where the local form of the
singularity becomes (dropping terms of higher than quadratic order)
\begin{equation}
-y^2 + 9tx-9t^2 + 9s^2 = (3s-y) (3s + y) + 9t (x-t) = 0
\end{equation}
There are two possible ways of blowing up this conifold singularity,
which we parameterize by $\tau \in\{0, 1\}$.  In each case we denote
the exceptional curve by $\tilde{\delta}_2$.
For $\tau = 0$ we
blow up the point at the origin into a $\P^1$ at $t = 3s-y = 0$,
parameterized by homogeneous coordinates $[x-t, 3s + y]$.
For $\tau = 0$ the curves $C_1^\pm$ intersect  $\tilde{\delta}_2$ at
the points
\begin{eqnarray}
C_1^+ \cap  \tilde{\delta}_2 & = &  \{\lim_{x, y, t, s \rightarrow
  0}[x, sx]
\}  =\{ [1, 0]\} \in \tilde{\delta}_2\\
C_1^-\cap \tilde{\delta}_2 & = &
\{   \lim_{x, y, t, s \rightarrow 0}[x,
  6s-sx]\}  = \{[x, 6s]\} = \tilde{\delta}_2
\end{eqnarray}
so for $\tau = 0$ the $s = 0$ limit of $C_1^-$ contains all of
$\tilde{\delta}_2$, while $C_1^+$, like $\delta_1$, intersects
$\tilde{\delta}_2$ at a single point.  A similar analysis for the
blow-up at $t = 3s + y = 0$ denoted by $\tau = 1$ shows that in this
case $C_1^+$ contains $\tilde{\delta}_2$, while $C_1^-$ intersects at
only a point.  So, just as in (\ref{eq:a3d4}), we can describe the
blow-up through a contribution of $\tau \tilde{\delta}_2$ to $C_2^+$
and $\bar{\tau} \tilde{\delta}_2$ to $C_2^-$, where $\bar{\tau} =
1-\tau$.
\vspace*{0.05in}

\noindent
{\bf Chart 2}: 

Blowing up the singularity at $(0, 0, 0, s)$ in (\ref{eq:a5-1})
in the $x$ chart gives
\begin{equation}
\Phi_{tx} = -y^2 + x^2 t +3 t x (t +1) 
 + s^2  (t +1)^2  = 0\,.
\label{eq:a5-2}
\end{equation}
The exceptional divisor at $x = 0$ on the threefold  is
\begin{equation}
C_2^\pm =\{(0, \pm s(t +1), t, s)\}
\end{equation}
for generic $s$, which degenerates to
\begin{equation}
\delta_2 =\{(0, 0, t, 0)\}
\end{equation}
at $s = 0$.
In this coordinate system, we have
\begin{equation}
C_1^\pm =\{(x, \pm s, 0, s)\} \,.
\end{equation}
There are singularities in (\ref{eq:a5-2}) at $(0, 0, -1, s)$ for all
$s$ and an additional singularity at $(0, 0, 0, 0)$ at $s = 0$.
The latter singularity is associated with the point $[1, 0, 0]$ in
homogeneous coordinates in the
$\P^2$ at the blown up point.  A similar analysis in the $t$ chart
shows an analogous singularity at $s = 0$ at the point $[0, 0, 1]$.

The isolated singularities at $s = 0$ take the conifold form and
can be resolved as above.  The singularity at
the origin in (\ref{eq:a5-2}) has the form
\begin{equation}
(s + y) (s-y) + 3tx = 0 \,.
\label{eq:conifold-double}
\end{equation}
Blowing up a $\P^1$ at $t = 0, s = \pm y$ gives a curve $\hat{\delta}_3$
that contributes to $C_1^\pm$ and to $C_2^\mp$.  The singularity at
homogeneous coordinates $[0, 0, 1]$ can be analyzed in the $t$ chart
and gives a curve $\check{\delta}_3$ that contributes only to $C_2^\pm$.
\vspace*{0.05in}

\noindent
{\bf Chart 3}: 

Finally, we can blow up the singularity at $(0, 0, -1, s)$ in
(\ref{eq:a5-2}) by shifting $t \rightarrow t-1$ and looking in the $x$
chart again, which gives
\begin{equation}
\Phi_{txx} = -y^2  +
   s^2 t^2 + 3 xt^2 + xt -3t-1= 0 \,.
\label{eq:a5-3}
\end{equation}
The exceptional divisor at $x = 0$ on the threefold  is the single curve
\begin{equation}
C_3 =\{(x, y, 0, s): y^2 = s^2 t^2 -3t -1\}
\end{equation}
for generic $s$, which degenerates to
\begin{equation}
\delta_3 =\{0, y,((y^2 +1)/3, 0)\}
\end{equation}
at $s = 0$.

By transforming the equations for each of the relevant curves into
each coordinate patch we see that the curves $\delta_{1, 2, 3},
\tilde{\delta}_2, \hat{\delta}_3,\check{\delta}_3$ have the correct
intersection matrix for $D_6$, with nonvanishing intersections
\begin{equation}
\tilde{\delta}_2 \cdot\delta_1  =
\delta_1 \cdot\hat{\delta}_3 =
\delta_2 \cdot\hat{\delta}_3 =
\delta_2 \cdot\check{\delta}_3 =
\delta_2 \cdot{\delta}_3 =
1\,.
\end{equation}
The curves $C_{1, 2}^\pm, C_3$ similarly
give $A_5$.  The embedding $A_5 \rightarrow D_6$ depends upon three
discrete parameters $\tau, \hat{\tau}, \tilde{\tau}$ describing the
choices for the conifold blow-ups, and is given by
\begin{eqnarray}
C_1^+ & \rightarrow &  \delta_1 + \tau \tilde{\delta}_2 + \hat{\tau}
\hat{\delta}_3 \nonumber\\
C_1^- & \rightarrow &  \delta_1 + \bar{\tau} \tilde{\delta}_2 + \bar{\hat{\tau}}
\hat{\delta}_3 \nonumber\\
C_2^+ & \rightarrow &  \delta_2 +  \bar{\hat{\tau}}
\hat{\delta}_3+  \check{\tau}
\check{\delta}_3\\
C_2^- & \rightarrow &  \delta_2 + \hat{\tau}
\hat{\delta}_3+  \bar{\check{\tau}}
\check{\delta}_3\nonumber\\
C_3 & \rightarrow &  \delta_3 \nonumber
\end{eqnarray}
It is straightforward to confirm that this embedding preserves all
inner products as needed.

From the decomposition of the adjoint of $D_5$, according to the
standard rank one reduction, we get a $\Lambda^2$ antisymmetric
representation of $SU(6)$.

\subsubsection{Enhancement $A_5 \subset E_6$}
\label{sec:a5-e6}

We now consider models
where $A_5$ is enhanced to $E_6$.  We begin with \eq{eq:phi-a5-e6}
\begin{equation}
\Phi = -y^2 + x^3 + \betah^2 x^2 +3 \betah x^2 t +
 3t^3x + 2 \betah t^2 x + t^4  = 0\,.
\label{eq:phi-a5-e6-appendix}
\end{equation}
As discussed in the main text, in (\ref{eq:phi-a5-e6-appendix}) the
parameter $\betah$ can be either $\betah = s$ or $\betah = s^2$.  As we
show below, for $\betah = s$ the $E_6$ singularity at $s = 0$ is not
completely resolved in the threefold, while it is in the case $\betah =
s^2$.  Most of the following analysis is independent of the power of
$s$ appearing in $\betah$.  We describe the differences in the
resolution for different choices of $\betah$ at the end of the
discussion.
\vspace*{0.05in}

\noindent
{\bf Chart 1}: 

Blowing up in the $t$ chart (\ref{eq:blow-up-t1}) gives
\begin{equation}
\Phi_t = -y^2 + x^3t + \betah^2 x^2 +3 \betah x^2 t +
 3t^2x + 2 \betah t x + t^2  = 0\,.
\label{eq:a5-e6-1}
\end{equation}
The exceptional divisor at $t = 0$ on the threefold  is
\begin{equation}
C_1^\pm =\{(x, \pm \betah x, 0, s)\}
\end{equation}
for generic $s$, which degenerates to
\begin{equation}
\epsilon_1 =\{(x, 0, 0, 0)\}
\end{equation}
at $s = 0$.
There is a singularity in (\ref{eq:a5-e6-1}) at $(0, 0, 0, s)$ for all
$s$.
\vspace*{0.05in}

\noindent
{\bf Chart 2}: 

Blowing up in the $x$ chart (\ref{eq:blow-up-x1}) gives
\begin{equation}
\Phi_{tx} = -y^2 + (\betah + t)^2 +
x^2 t  +3 \betah x t +
 3t^2x= 0\,.
\label{eq:a5-e6-2}
\end{equation}
The exceptional divisor at $x = 0$ on the threefold  is
\begin{equation}
C_2^\pm =\{(0, \pm  (\betah + t), t, s)\}
\end{equation}
for generic $s$, which degenerates to
\begin{equation}
\epsilon_2^\pm =\{(0, \pm t, t, 0)\}
\end{equation}
at $s = 0$.
There is a singularity in (\ref{eq:a5-e6-2}) where $t = -\betah$, at
$(0, 0,  -\betah, s)$ for all $s$.
\vspace*{0.05in}

\noindent
{\bf Chart 3}: 

Shifting $t \rightarrow t-\betah$ and
blowing up again in the $x$ chart (\ref{eq:blow-up-x1}) gives
\begin{equation}
\Phi_{txx} = -y^2 + t^2 -3 \betah t -\betah + xt
+ 3t^2x= 0\,.
\label{eq:a5-e6-3}
\end{equation}
The exceptional divisor at $x = 0$ on the threefold  is
\begin{equation}
C_3 =\{(0, y, t, s):y^2 = t^2 -3 \betah t-\betah\}
\end{equation}
for generic $s$, which degenerates to
\begin{equation}
\epsilon_3^\pm =\{(0, \pm t, t, 0)\}
\end{equation}
at $s = 0$.

At this point, the structure of the singularity
at $(0, 0, 0, 0)$ in chart 3
depends upon whether
$\betah = s$ or $\betah = s^2$.  In either case, on the slice $s = \betah
= 0$ (\ref{eq:a5-e6-3}) has a singularity at the origin.  If, however,
$\betah = s$, then in the full space there is no singularity, and no
further resolution is necessary.  In this case the $E_6$ singularity
is not completely resolved, and the 3 curves $\epsilon_1,
\epsilon_3^\pm$ all intersect at a point.  If, on the other hand,
$\betah = s^2$ then we have another conifold type singularity at the
origin
\begin{equation}
(is + y) (is-y) + t (x + t) \,.
\end{equation}
Resolving this  singularity in either way gives another
curve $\epsilon_4$, which completes the resolution of $E_6$.

As above, the intersections of the various curves can be worked out to
give the explicit embedding of $A_5$.  
Note that $C_2^{\pm}, \epsilon_2^\pm$ are not visible in chart 3; to see
these and their intersections with $C_3, \epsilon_3^\pm$, another
({\it e.g.}, $t$) coordinate patch is needed.
In the case of the incompletely
resolved $E_6$ when $\betah = s$, the embedding is
\begin{eqnarray}
C_1^\pm  & \rightarrow &  \epsilon_1 + \epsilon_3^\pm \nonumber\\
C_2^\pm  & \rightarrow & \epsilon_2^\pm \label{eq:c-e}\\
C_3 & \rightarrow &  \epsilon_3^+ + \epsilon_3^- \nonumber
\end{eqnarray}
This embedding is depicted graphically in Figure~\ref{f:a5-e6x}.

For $\betah = s^2$, the $E_6$ at $s = 0$ is completely resolved, and
the embedding is
\begin{eqnarray}
C_1^\pm  & \rightarrow &  \epsilon_1 + \epsilon_4 + \epsilon_3^\pm \nonumber\\
C_2^\pm  & \rightarrow & \epsilon_2^\pm\\
C_3 & \rightarrow &  \epsilon_3^+ + \epsilon_3^-+ \epsilon_4 \nonumber
\end{eqnarray}
This embedding is depicted graphically in Figure~\ref{f:a5-e6}.

In the case $\betah = s$ the matter content contains
a half hypermultiplet in the 3-index antisymmetric $\Lambda^3$
representation, while for $\betah = s^2$ there is a full hypermultiplet
in this representation, as discussed in the main text.

\subsection{Enhancement of $A_3 \rightarrow A_7$ at an ordinary double
  point}
\label{sec:ordinary-double-appendix}

Now we consider a situation where the curve $\{\sigma=0\}$ itself becomes
singular.  As discussed in Section \ref{sec:double}, in such a
situation there will be a matter representation with nonzero genus
contribution.  
We consider the ordinary double point singularity in
eq.\ (\ref{eq:ordinary-double})
\begin{equation}
\Phi =
-y^2 + x^3 + x^2 -s^2 t^2 x = 0 \,.
\end{equation}
This gives a Calabi-Yau threefold with an $A_3$  singularity along the lines
$s = 0$ and $t = 0$ with an enhancement to $A_7$ at the point $s = t =
0$.  We can resolve the singularity by first resolving the $t = 0$
singularity in a sequence of charts 1, 2 identical to those used in
\ref{sec:a3-resolution}.  In the first chart we have $C_1^\pm$ as
before, which correspond to curves $\gamma_1^\pm$ in the $A_7$
resolution.  In the
second chart, using (\ref{eq:blow-up-t1}) we have
\begin{equation}
\Phi_{tt} = -y^2 + x^3t^2 + x^2 -s^2 x \,,
\end{equation}
with exceptional curve
\begin{equation}
C_2 =\{(x, y, 0, s): y^2 = x^2 -s^2 x\}
\end{equation}
that at $s = 0$ becomes
\begin{equation}
\gamma_2^\pm =\{(x, \pm x, 0, s)\} \,.
\end{equation}
\vspace*{0.05in}

\noindent
{\bf Chart 3}: 

Now we blow up the singularity at the origin again using the
coordinate transformation
\begin{equation}
(x, y, t, s)_2 = (xs, ys, t, s)_3 \,.
\label{eq:blow-up-s1}
\end{equation}
This gives (dropping the $x^3$ term that is irrelevant for the analysis)
\begin{equation}
\Phi_{tts} = -y^2 +x^2 -s x \,.
\end{equation}
This has exceptional curves at $s = 0$ given by
\begin{equation}
\tilde{C}_1^\pm =\{(x, \pm x, 0, 0)\},
\end{equation}
which we identify with curves $\gamma_3^\pm$.
\vspace*{0.05in}

\noindent
{\bf Chart 4}: 

Blowing up one more time using (\ref{eq:blow-up-s1}) gives
\begin{equation}
\Phi_{ttss} =-y^2 +x^2 -x \,.
\end{equation}
This defines a nonsingular curve $y^2 = x^2 -x$, which we identify as
$\tilde{C}_2 = \gamma_4$.

Following the coordinate charts and determining the intersections of
the various curves we have
\begin{eqnarray}
C_1^\pm & \rightarrow &  \gamma_1^\pm\\
C_2 & \rightarrow &  \gamma_2^+ + \gamma_2^-+ \gamma_3^+ + \gamma_3^-+ \gamma_4\\
\tilde{C}_1^\pm & \rightarrow &  \gamma_3^\pm\\
\tilde{C}_2 & \rightarrow & \gamma_4
\end{eqnarray}
Depending upon how $A_3$ is embedded into the $\tilde{C}$'s relative
to the $C$'s,
this gives an embedding $SU(4)  \rightarrow SU(4) \times SU(4)
\rightarrow SU(8)$ under which the decomposition of the adjoint
can include either an extra adjoint field or
includes a symmetric (Sym${}^2$) and an antisymmetric ($\Lambda^2$)
representation.
As illustrated in the main text, if the root of $A_3$ associated with
$C_1^+$ is also associated with $\tilde{C}_1^-$ then the
representation content is symmetric plus antisymmetric.
\vspace*{0.1in}

{\bf Acknowledgements}: We would like to thank Volker Braun, Antonella Grassi,
Thomas Grimm, Sheldon Katz, Vijay
Kumar, Joe Marsano, Daniel Park, James Sully, Sakura Sch\"afer-Nameki,
and Cumrun Vafa for helpful
discussions.  Thanks to the Institute for Physics and Mathematics of
the Universe (IPMU) for hospitality during the initial phases of this
work.  WT would like to thank UCSB and DRM would like to thank MIT for
hospitality in the course of this work.
This research was supported by the DOE under contract
\#DE-FC02-94ER40818, by the National Science Foundation under grant
DMS-1007414, and by World Premier International Research Center Initiative (WPI Initiative), MEXT, Japan.

\end{document}